\documentclass[a4paper,10pt]{article}

\usepackage{amssymb}
\usepackage{amsmath}
\usepackage{subfigure}
\usepackage{slashed}

\allowdisplaybreaks

\usepackage{soul}

\usepackage{graphicx}
\usepackage{amsthm}
\usepackage{latexsym}
\usepackage[dvips]{epsfig}

\usepackage{wasysym}
\usepackage{mathrsfs}
\usepackage{eufrak}
\usepackage{bm}
\usepackage{authblk}
\usepackage{slashed}
\usepackage{yhmath} 
\usepackage{stmaryrd}

\theoremstyle{plain}
\newtheorem{proposition}{Proposition}

\newtheorem{assumption}{Assumption}

\newtheorem{remark}{Remark}

\def\bma{{\bm a}}
\def\bmb{{\bm b}}
\def\bmc{{\bm c}}

\def\bme{{\bm e}}

\def\bmg{{\bm g}}

\def\bmA{{\bm A}}
\def\bmB{{\bm B}}
\def\bmC{{\bm C}}

\def\bmeta{{\bm \eta}}

\def\bmsigma{{\bm \sigma}}

\newcounter{mnotecount}

\newcommand{\mnotex}[1]
{\protect{\stepcounter{mnotecount}}$^{\mbox{\footnotesize $\bullet$\themnotecount}}$ 
\marginpar{
\raggedright\tiny\em
$\!\!\!\!\!\!\,\bullet$\themnotecount: #1} }

\newcommand{\bit}{\begin{itemize}}
\newcommand{\eit}{\end{itemize}}
\newcommand{\ben}{\begin{enumerate}}
\newcommand{\een}{\end{enumerate}}

\newcommand{\beq}{\begin{equation}}
\newcommand{\eeq}{\end{equation}}
\newcommand{\bea}{\begin{eqnarray}}
\newcommand{\eea}{\end{eqnarray}}
\newcommand{\nn}{\nonumber}

\newcommand{\bwt}{\begin{widetext}}
\newcommand{\ewt}{\end{widetext}}

\usepackage[utf8]{inputenc}
\usepackage[english]{babel}
 
\usepackage{multicol}
\usepackage{color}
 
\usepackage{comment}
 
\begin{document}

\title{\textbf{Spectral methods for the spin-2 equation near the
    cylinder at spatial infinity }}

\author[,1]{Rodrigo P. Macedo \footnote{E-mail address:{\tt
      r.panossomacedo@qmul.ac.uk}}} \author[,1]{Juan A. Valiente
  Kroon \footnote{E-mail address:{\tt j.a.valiente-kroon@qmul.ac.uk}}}
\affil[1]{School of Mathematical Sciences, Queen Mary, University of
  London, Mile End Road, London E1 4NS, United Kingdom.}

\maketitle

\begin{abstract}
We solve, numerically, the massless
spin-2 equations, written in terms of a gauge based on the properties
of conformal geodesics, in a neighbourhood
of spatial infinity using spectral methods in both space and
time. This strategy allows us to compute the solutions to these
equations up to the critical sets where null infinity intersects with
spatial infinity. Moreover, we use the convergence rates of the
numerical solutions to read-off their regularity properties.
\end{abstract}



\section{Introduction}

The conformal Einstein field equations introduced by Friedrich
\cite{Fri81a,Fri84} ---see also \cite{CFEBook}--- provide a powerful
tool for the study of the global properties of spacetimes. They also
provide a natural framework for the numerical construction of global
solutions to the Einstein field equations ---see
e.g. \cite{Hub99a,Hub99b,Hub01b}. In this context, of
particular interest are solutions which can be described as
\emph{Minkowski-like} ---i.e. vacuum spacetimes admitting a conformal
(Penrose) extension with the same qualitative properties of the
standard compactification of the Minkowski spacetime, see
e.g. \cite{CFEBook} chapter 6. In particular, the numerical simulations reported in
\cite{Hub01b} provide an illustration of the semi-global stability
result of the Minkowski spacetime from a hyperboloidal initial value
problem ---see \cite{Fri86b,LueVal09}. The extension of Friedrich's
semiglobal  existence results to a true global stability result
depends on the resolution of the so-called \emph{problem of spatial
  infinity}  ---i.e. the development of analytical methods to deal
with the singular behaviour of the conformal structure of spacetime at
spatial infinity. For a discussion of the background and context of
this see e.g. \cite{CFEBook}, chapter 20. 

At the core of the conformal Einstein field equations lies the
\emph{spin-2
equation} satisfied by the so-called \emph{rescaled Weyl tensor}. The central role of
this subsystem is better brought to the foreground in a gauge based on
the properties of certain conformal invariants ---the so-called
\emph{conformal geodesics}, see e.g. \cite{Fri98a,Fri03a,CFEBook}. In the
following we will refer to this gauge as the \emph{F-gauge}. In this
gauge it is possible to derive a system of evolution equations in
which all the conformal fields, except for the rescaled Weyl tensor
satisfy transport equations (i.e. ordinary differential equations)
along the conformal geodesics. The only true partial differential
equations in this hyperbolic reduction arises from the Bianchi
equations for the Weyl tensor. From the above discussion it follows that a convenient
model problem to study the properties of the conformal field
equations near spatial infinity is the analysis of the propagation of \emph{massless spin-2
fields} on a (fixed) Minkowski background. 

One of the central features of the F-gauge used in the seminal
study of the problem of spatial infinity in \cite{Fri98a} is
that it gives rise to a representation of spatial infinity in which the
point $i^0$ is blown-up to a cylinder ---the \emph{cylinder at spatial
  infinity}. This cylinder can be identified with the \emph{spatial
  infinity hyperboloid} discussed in studies of the conformal
structure of spacetime in the 1970's and 1980's ---see
e.g. \cite{AshHan78,Ash80,BeiSch82,Bei84}. Crucially, the cylinder at
spatial infinity is a \emph{total characteristic} of the conformal
evolution equations ---that is, the whole evolution equations reduce
to a system of total characteristics at this part of the conformal
boundary. This remarkable interplay between the (conformal) geometry
and the structural properties of the evolution equations allows us to 
resolve with great detail the (generic) singular behaviour of the solutions of the
conformal evolution equations as one approaches the sets where the
spatial infinity and null infinity intersect. 

A systematic analytical discussion of the massless spin-2 equation propagating in a
neighbourhood of the spatial infinity of the Minkowski spacetime has been
given in \cite{Val03a,Fri03b} ---see also
\cite{GasVal16}. The properties of the solutions of the conformal evolution
equations constitute a symmetric hyperbolic system which degenerates
at the \emph{critical sets} where spatial infinity and null infinity
meet. More precisely the matrix associated to the time derivatives of
the fields, which normally should be positive definite, looses
rank. Accordingly, the standard analytic methods to control the
solutions of hyperbolic equations do not apply at these
sets. This observation dominates the properties of the solutions. In
particular, generic solutions will develop logarithmic singularities
at the critical sets. These singular behaviour can be avoided if the
initial data is fine-tuned in a particular manner. 

\medskip
As a result  of the degeneracy of the evolution equations at the
critical points, the numerical evaluation of the solutions to the
massless spin-2 equations near spatial infinity (and more generally,
the full Einstein field equations) poses particular challenges. 

Numerical studies of the spin-2 equations in a neighbourhood of spatial infinity were
first presented in \cite{BeyDouFraWha12,BeyDouFraWha13} (first order
formulation) and \cite{DouFra13} (second order
formulation). While the focus was originally on the behaviour of the
fields around $i^0$, global evolutions were discussed more recently in
\cite{DouFra16}.  These studies provide valuable insight and
intuition into the advantages and difficulties of different
representations of the critical sets and null infinity. Moreover, they
examine the limitations to resolve those regions with
mainstream numerical algorithms, such as the explicit time integrator
Runge Kutta 4 and suggest that a better way of carrying out these
numerical evaluations is by means of \emph{spectral methods}. 

Spectral methods are a well stablished tool to solve elliptic
equations ---see \cite{CanHusQuaZan06} for a classical textbook and
\cite{GraNov07} for applications in General Relativity.  Over the last
decade, M. Ansorg fostered the idea of extending the applicability of
spectral methods and include the time direction as well.  In this
respect, fully (pseudo-)spectral methods --- where the spectral
decomposition is applied to {\em both} space and time directions ---
have been adapted to the solution of hyperbolic equations \cite{HenAns09,MacAns14}.  In particular, fully
spectral codes have been used for the study of scalar fields around
the spatial infinity either on the Minkowski background
\cite{FraHen14} or on the Schwarzschild spacetime
\cite{FraHen17,FraHen18}. Further applications can be found in \cite{AnsHen11, Hen12, AmmGriJimMacMel16,SanSop18}.

\medskip
In this article we investigate the possibility of solving the massless
spin-2 equations, written in terms of the F-gauge, in a neighbourhood
of spatial infinity using spectral methods in both space and
time. Moreover, we also address the question of \emph{whether it is
possible to resolve the regularity properties of the numerical
solutions by direct inspection of the numerical error convergence
rates}. As it will be seen in the main text we answer this question in
the positive. Spectral methods in time are a robust implicit method
allowing us to deal with the troublesome critical sets in a
straightforward way. Analytic (i.e. entire) solutions display an
exponential convergence rate of the numerical error. By contrast,
solutions with logarithmic singularities spoil the method's fast
convergence rate. Although in this case the numerical error decays at
a merely algebraic rate, we can use this feature on our advantage to
scrutinise the underlying regularity of the solution.

\subsection{Outline of the paper}
In section \ref{CylinderAtspatialInfinity} we review a conformal
representation of the Minkowski spacetime which is suitable for
studying the behaviour of fields near spatial infinity, while in
section \ref{LinearisedGravityEquations} we perform the analysis of a
spin-2 field propagating on this spacetime. In section
\ref{NumericalSolution} the equations are adapted to the numerical
implementation. In particular, we discuss the numerical methods
employed in the solution and show the results. Finally, we present the
discussion and conclusion in section \ref{DiscussionConclusion}.

\subsection{Notations and Conventions}

 The signature convention for (Lorentzian) spacetime metrics will be $
 (+,-,-,-)$.  In this article $\{_a ,_b , _c ,
 . . .\}$ denote abstract tensor indices and $\{_\bma ,_\bmb , _\bmc ,
 . . .\}$ will be used as spacetime frame indices taking the values ${
   0, . . . , 3 }$.  In this way, given a basis
$\{\bme_{\bma}\}$ a generic tensor is denoted by $T_{ab}$ while its
components in the given basis are denoted by $T_{\bma \bmb}\equiv
T_{ab}\bme_{\bma}{}^{a}\bme_{\bmb}{}^{b}$.
 Part of  the analysis will require the use of spinors.
In this respect, the notation and
 conventions of Penrose \& Rindler \cite{PenRin84} will be followed.
 In particular, capital Latin indices $\{ _A , _B , _C , . . .\}$ will
 denote abstract spinor indices while boldface capital Latin indices
 $\{ _\bmA , _\bmB , _\bmC , . . .\}$ will denote frame spinorial
 indices with respect to  a specified spin dyad ${
   \{\delta_\bmA{}^{A} \} }.$
The  conventions for the curvature tensors are fixed by the relation
\[
(\nabla_a \nabla_b -\nabla_b \nabla_a) v^c = R^c{}_{dab} v^d.
\]

\section{The cylinder at spatial infinity and the F-Gauge}
\label{CylinderAtspatialInfinity}

In this section we discuss a conformal representation of the Minkowski spacetime
adapted to a congruence of conformal geodesic. This conformal
representation was introduced in
\cite{Fri98a} and is particularly suited for the analysis of the behaviour of
fields near spatial infinity. Roughly speaking, in this representation
spatial infinity $i^{0}$, which corresponds to a point in the standard
compactification of the Minkowski spacetime, is blown up to a
2-sphere $\mathbb{S}^2$ ---this representation is called  the
\emph{cylinder at spatial infinity}. The original 
discussion of the cylinder at spatial infinity as presented in
\cite{Fri98a} is given in the language of fibre bundles. A
presentation of this construction which does not make use of this
language was given in \cite{GasVal16}. Here we follow this
presentation. 

\subsection{The cylinder at spatial infinity}
\label{TheCylinderAtSpatialInfinity}
\begin{figure*}[t!]
\begin{center}
\includegraphics[width=7.1cm]{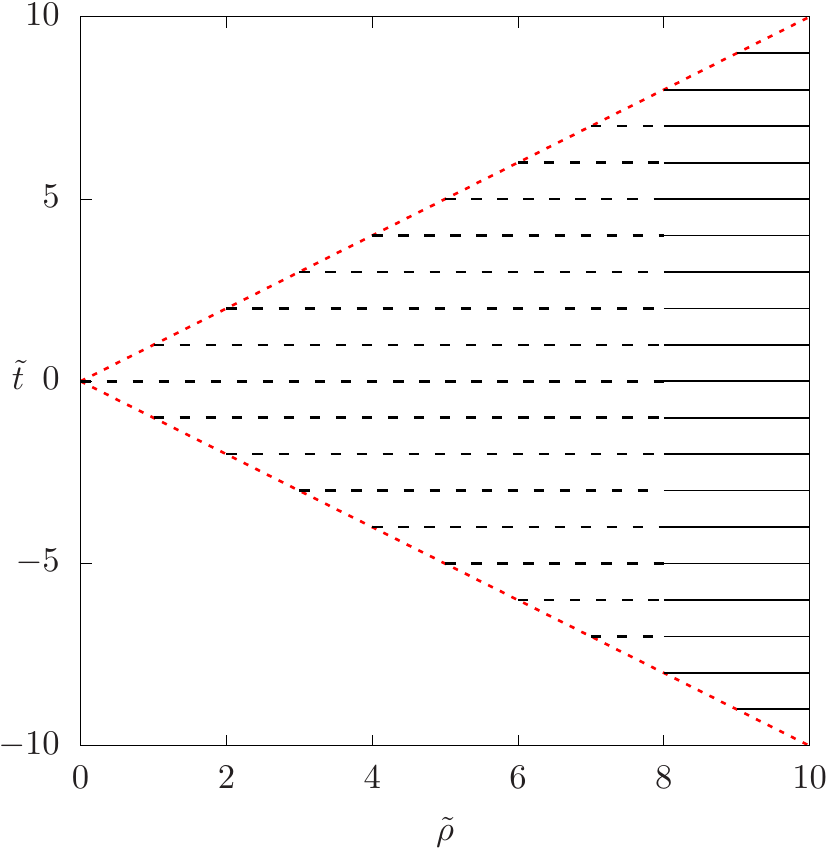}
\includegraphics[width=7.4cm]{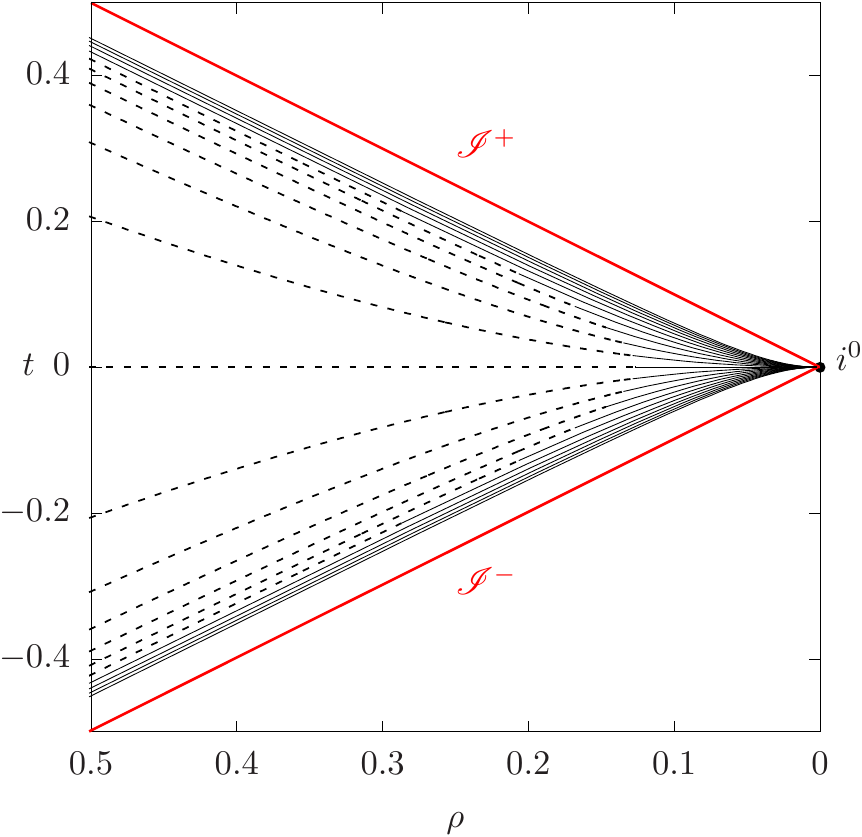}
\end{center}
\caption{Conformal map for the Minkowski spacetime via the inversion
coordinates. Left panel: polar coordinates $(\tilde{t}, \tilde\rho)$
for the physical spacetime with metric $\boldsymbol{\tilde\eta}$. The
red dotted lines correspond to the light cones given by $\tilde{X} =
0$. The surfaces $\tilde{t}=$ constant are drawn in black within the
region $\tilde{X}<0$. Dashed lines represent the region near the
symmetry axis $\tilde{\rho}=0$, whereas straight lines correspond to
their prolongation towards $\tilde\rho\rightarrow\infty$. Right panel:
polar coordinates $(t, \rho)$ for the unphysical conformal spacetime
with metric $\boldsymbol{g}$. The surfaces $\tilde{t}=$ constant are
represented in the new coordinates. The straight black lines
representing the asymptotic region $\tilde\rho\rightarrow\infty$
converge at the point $(0,0)$ ---i.e., at spacelike infinity $i^0$.
}
\label{fig:InversionCoord}
\end{figure*}

To start, consider the Minkowski metric $\tilde{\bmeta}$ expressed in Cartesian
coordinates $(\tilde{x}^{\alpha })= (\tilde{t},\tilde{x}^{i})$,
 \[
 \tilde{\bmeta}=\eta_{\mu\nu}\mathbf{d}\tilde{x}^{\mu}\otimes\mathbf{d}\tilde{x}^{\nu},
 \]
 where $\eta_{\mu\nu}=\text{diag}(1,-1,-1,-1)$. By introducing polar
 coordinates defined by $\tilde{\rho}^2\equiv 
 \delta_{ij}\tilde{x}^{i}\tilde{x}^{j}$ where
 $\delta_{ij}=\text{diag(1,1,1)}$, and an arbitrary choice of
 coordinates on $\mathbb{S}^2$, the metric $\tilde{\bmeta}$ can be
 written as
\begin{equation*}
\tilde{\bmeta}=\mathbf{d}\tilde{t}\otimes\mathbf{d}\tilde{t}
-\mathbf{d}\tilde{\rho}\otimes \mathbf{d}\tilde{\rho}-\tilde{\rho}^2
\mathbf{\bm\sigma},
\end{equation*}
with $\tilde{t}\in(-\infty, \infty)$, $\tilde{\rho}\in [0,\infty)$ and where
  $\bm\sigma$ denotes the standard metric on $\mathbb{S}^2$.  A
  strategy to construct a conformal representation of the Minkowski
  spacetime close to $i^{0}$ is to make use of  \emph{inversion coordinates} $(x^{\alpha})=(t,x^{i})$ defined by ---see
  \cite{Ste91}---
 \[ 
x^{\mu}=-{\tilde{x}^{\mu}}/{\tilde{X}^2}, \qquad \tilde{X}^2 \equiv
 \tilde{\eta}_{\mu\nu}\tilde{x}^{\mu}\tilde{x}^{\nu}.
\]
 The inverse transformation is given by
 \[
\tilde{x}^{\mu}=-x^{\mu}/X^2, \qquad
 X^2=\eta_{\mu\nu}x^{\mu}x^{\nu}.
\]
Observe, in particular that $X^2=1/\tilde{X}^2$. Using these
coordinates one identifies a conformal representation of the Minkowski
spacetime with \emph{unphysical metric} given by
\[
\bmg=\Xi^2  \tilde{\bmeta},
\]
where $\bmg=\eta_{\mu\nu}\mathbf{d}x^{\mu}\otimes \mathbf{d}x^{\nu}$
and $\Xi =X^2$. Introducing an \emph{unphysical polar coordinate} via
the relation $\rho^2\equiv
\delta_{ij}x^{i}x^{j}$, one finds that the metric $\bmg$ can be
written as
\[
\bmg=\mathbf{d}t\otimes\mathbf{d}t -\mathbf{d}\rho\otimes
\mathbf{d}\rho-\rho^2 \mathbf{\bm\sigma}, \qquad \Xi=t^2-\rho^2,
\]
with $t\in(-\infty,\infty)$ and $\rho\in (0,\infty)$.
In this conformal representation, spatial infinity $i^{0}$ corresponds
to  a point located at the origin --- see Fig.~\ref{fig:InversionCoord}. Observe that $\tilde{t}$ and $\tilde{\rho}$
are related to $t$ and $\rho$ via
\[
\tilde{t}=-\dfrac{t}{t^2-\rho^2}, \qquad  \tilde{\rho}= -\dfrac{\rho}{t^2-\rho^2}.
\]

 Finally, introducing a time coordinate
$\tau$ through the relation  $t=\rho\tau$ one finds that the metric
$\bmg$ can be written as
\begin{equation*}
\bmg=\rho^2 \mathbf{d}\tau\otimes \mathbf{d}\tau
-(1-\tau^2)\mathbf{d}\rho \otimes \mathbf{d}\rho + \rho\tau
\mathbf{d}\rho\otimes \mathbf{d}\tau + \rho\tau \mathbf{d}\tau \otimes
\mathbf{d}\rho - \rho^2 \bmsigma.
\end{equation*}
The conformal representation containing the \emph{cylinder at spatial infinity} is obtained by considering the rescaled metric 
\[
\bar{\bmg} \equiv \dfrac{1}{\rho^2} \bmg.
\]
Introducing the coordinate  $\varrho\equiv-\ln \rho$ the metric $\bar{\bmg}$
can be reexpressed as 
\[
\bar{\bmg}=\mathbf{d}\tau \otimes \mathbf{d}\tau
-(1-\tau^2)\mathbf{d}\varrho \otimes \mathbf{d}\varrho
- \tau \mathbf{d}\tau \otimes \mathbf{d}\varrho -\tau
\mathbf{d}\varrho \otimes \mathbf{d}\tau -\bm\sigma.
\]
Observe that  spatial infinity $i^{0}$, which is at infinity 
 respect to the metric $\bar{\bmg}$, corresponds to a set
which has the topology of $\mathbb{R}\times\mathbb{S}^2$ ---see
\cite{Fri98a, AceVal11}.  Following the previous discussion, 
 one considers the conformal extension $(\mathcal{M},\bar{\bmg})$ where
\[
\bar{\bmg}=\Theta^2 \tilde{\bm\eta}, \qquad \Theta=\rho(1-\tau^2),
\]
and 
\[
\mathcal{M} \equiv \big\{ p \in \mathbb{R}^4 \; \rvert \; -1 \leq \tau \leq 1 , \; \; \rho(p)\geq 0\big\}.
\]
In this representation future and past null infinity are described by
the sets
\[
 \mathscr{I}^{+} \equiv \big\{ p \in \mathcal{M} \; \rvert\; \tau(p) =1 \big\}, 
\qquad \mathscr{I}^{-} \equiv \big\{ p \in \mathcal{M} \; \rvert \; 
 \tau(p) =-1\big\},
\]
while the physical Minkowski spacetime can be
identified with the set
\[
\tilde{\mathcal{M}} \equiv \big\{ p \in \mathcal{M} \; \rvert \; -1<\tau(p)<1 , \; \;\rho(p)>0 \big\},
\]
In addition, the following sets play a role in our discussion:
\begin{equation*}
 I \equiv \big\{ p \in \mathcal{M} \; \rvert   \;\;  |\tau(p)|<1, \; \rho(p)=0\big\}, 
\qquad I^{0} \equiv \big\{ p \in \mathcal{M}\; \rvert \;
  \tau(p)=0, \; \rho(p)=0\big\},
\end{equation*} 
and 
\begin{equation*}
 I^{+} \equiv \big\{ p\in \mathcal{M} \; \rvert \; \tau(p)=1, \; \rho(p)=0
  \big\}, \qquad I^{-} \equiv \big\{p \in \mathcal{M}\; \rvert \; \tau(p)=-1, \; \rho(p)=0\big\}.
\end{equation*}

 \begin{figure*}[b!]
\begin{center}
\includegraphics[width=7.1cm]{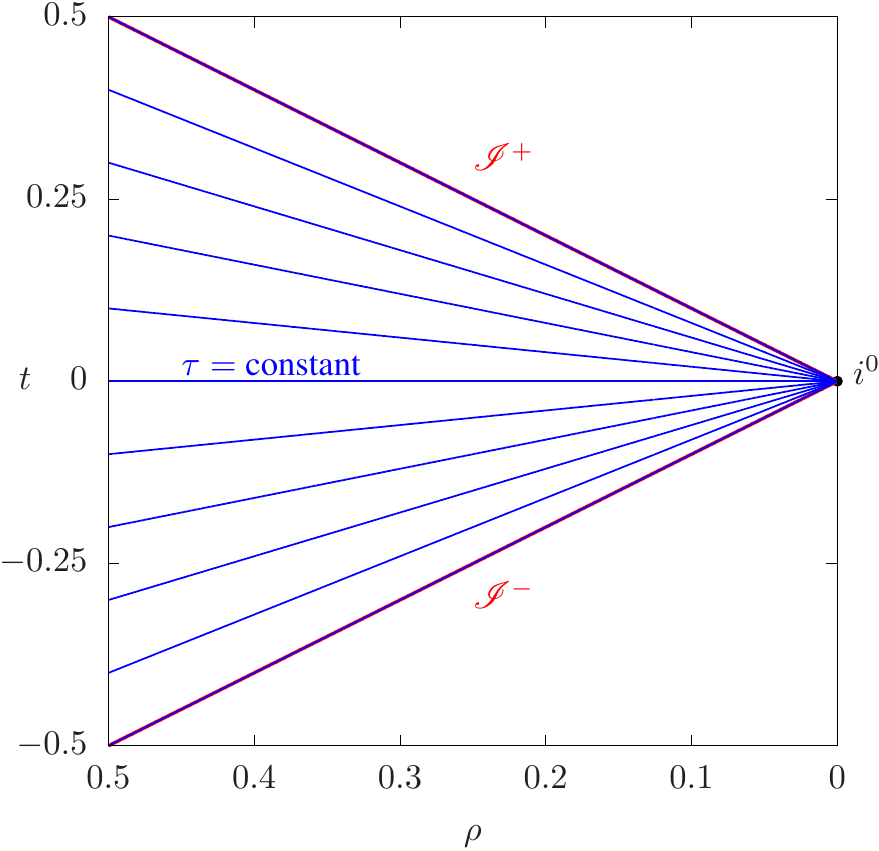}
\includegraphics[width=7.5cm]{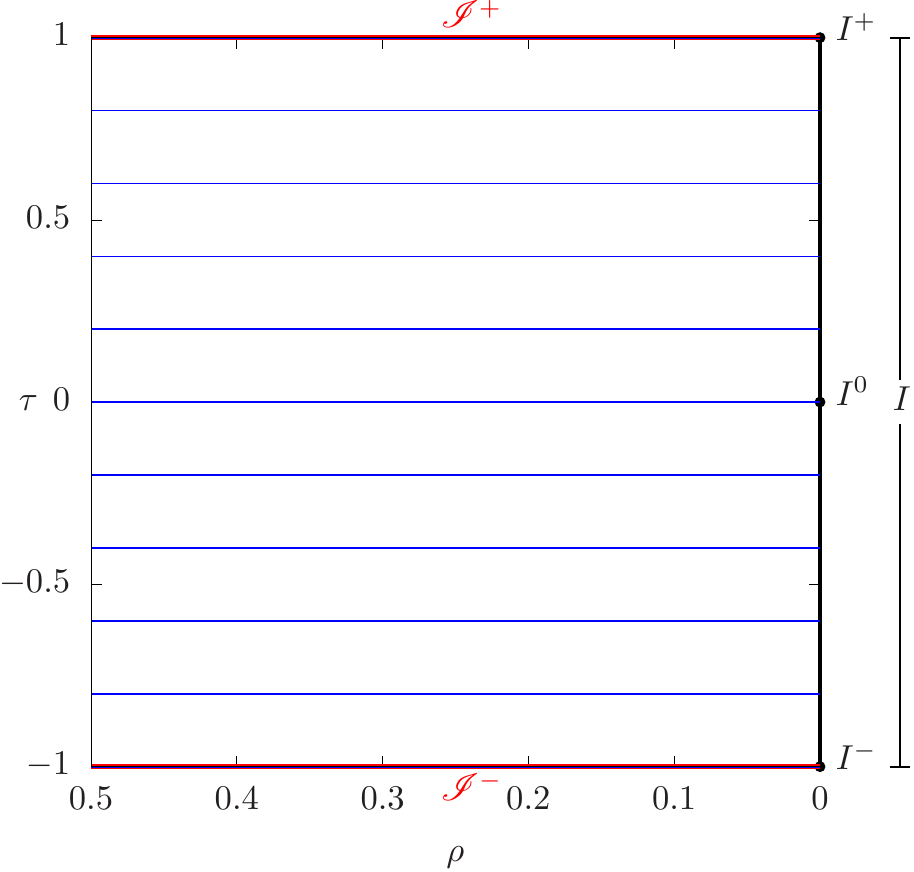}
\end{center}
\caption{
Cylinder at spatial infinity. Left panel: unphysical conformal
spacetime with metric $\boldsymbol{g}$ in polar coordinates $(t,
\rho)$. The straight blue lines correspond to hypersurfaces with
constant new time coordinate. Right panel: unphysical conformal
spacetime with metric $\boldsymbol{\bar{g}}$ in coordinates $(\tau,
\rho)$. The cylinder at space infinity is represented by the set
$I$. The critical sets $I^\pm$ are the points where spatial infinity
intersects null infinities $\mathscr{I}^\pm$ while $I^0$ corresponds
to the intersection of $i^0$ with $\tau=0$.
}
\label{fig:Cylinder}
\end{figure*}

\begin{remark}
{\em Observe in Fig.~\ref{fig:Cylinder} that spatial infinity $i^0$, a point in the
$\bmg-$representation, is identified with the set $I$ in the
$\bar{\bmg}-$representation. The \emph{critical sets} $I^{+}$ and
$I^{-}$ are the collection of points where 
spatial infinity intersects, respectively, $\mathscr{I}^{+}$ and $\mathscr{I}^{-}$. Similarly, $I^{0}$ is the intersection of
$i^{0}$ and the initial hypersurface $\mathcal{S} \equiv
\{\tau=0\}$. See \cite{Fri98a, FriKan00} and \cite{AceVal11} for
further discussion of the framework of the cylinder at spatial in stationary spacetimes.}
\end{remark}

\subsection{The F-gauge}
\label{SectionF-gauge}

In this section we provide a brief discussion of the so-called
\emph{F-gauge} ---see \cite{FriKan00,AceVal11} for a discussion of the
F-gauge in the language of fibre bundles. One of the chief
motivations for the introduction of this gauge is that it is based on
the properties of conformal geodesics. Accordingly, one introduces a
null frame whose timelike leg corresponds to the tangent of a
conformal geodesic starting from a fiduciary spacelike hypersurface
$\mathcal{S}=\{\tau=0\}$. For a discussion of the properties of
conformal geodesics see e.g. \cite{Fri95,Fri03c,Tod02,CFEBook}.

\medskip
 Consider the conformal extension
$(\mathcal{M},\bar{\bmg})$ of the Minkowski spacetime and the
F-coordinate system introduced in Section
\ref{TheCylinderAtSpatialInfinity}. The induced metric on the 2-sphere $ \mathcal{Q}\equiv\{
\tau=\tau_{\star},\rho=\rho_{\star}, \}$, with $\tau_{\star},
\rho_{\star}$ fixed, is the standard metric on $\mathbb{S}^2$. On
these 2-spheres one can 
introduce a complex null frame $\{ \bm\partial_{+},\bm\partial_{-}\}$
on $\mathcal{Q}$. This frame is propagated 
$\mathcal{Q}$ by requiring the conditions 
\[
[\bm\partial_{\tau},\bm\partial_{\pm}]=0, \qquad
[\bm\partial_{\rho},\bm\partial_{\pm}]=0.
\]
The above vector fields are used to define the spacetime frame
\begin{align*}
&\bme_{\bm0\bm0'}=\dfrac{\sqrt{2}}{2}\big((1-\tau)\bm\partial_{\tau} +
\rho\bm\partial_{\rho}\big), \qquad 
\bme_{\bm1\bm1'}=\dfrac{\sqrt{2}}{2}\big((1+\tau)\bm\partial_{\tau} -
\rho\bm\partial_{\rho}\big), \\
&\bme_{\bm0\bm1'}=\dfrac{\sqrt{2}}{2}\bm\partial_{+}, \qquad \qquad
 \qquad \qquad \quad
\bme_{\bm1\bm0'}=\dfrac{\sqrt{2}}{2}\bm\partial_{-}. 
\end{align*}
The corresponding dual coframe is given by
\begin{align*}
&\bm\omega^{\bm0\bm0'}=\dfrac{\sqrt{2}}{2}\Big( \mathbf{d}\tau
-\dfrac{1}{\rho}\big(1-\tau\big)\mathbf{d}\rho\Big),\qquad
\bm\omega^{\bm1\bm1'}=\dfrac{\sqrt{2}}{2}\Big(\mathbf{d}\tau +
\dfrac{1}{\rho}\big(1+\tau\big)\mathbf{d}\rho\Big),\\
&\bm\omega^{\bm0\bm1'}=\sqrt{2}\bm\omega^{+}, \qquad \qquad \qquad \qquad 
\qquad \quad \bm\omega^{\bm1\bm0'}=\sqrt{2}\bm\omega^{-},
\end{align*}
with $\boldsymbol{\omega}^\pm$ dual to $\boldsymbol{\partial}_\pm$
---i.e. one has the pairings
\[
\langle \boldsymbol{\omega}^+, \boldsymbol{\partial}_+ \rangle =1,
\qquad \langle \boldsymbol{\omega}^-, \boldsymbol{\partial}_- \rangle
=1, \qquad \langle \boldsymbol{\omega}^+, \boldsymbol{\partial}_-
\rangle =0, \qquad \langle \boldsymbol{\omega}^-, \boldsymbol{\partial}_+ \rangle =0.
\]
In terms of some polar coordinates $(\theta,\varphi)$ on $\mathbb{S}^2$
one can write 
\[
\boldsymbol{\partial}_\pm =  \frac{1}{\sqrt{2}}\bigg(\boldsymbol{\partial}_\theta \pm
\frac{\mbox{i}}{\sin\theta}\boldsymbol{\partial}_\varphi\bigg)
\qquad \boldsymbol{\omega}^\pm = \frac{1}{\sqrt{2}}\big(\mathbf{d}\theta
\mp \mbox{i}\sin\theta \mathbf{d}\varphi
\big).
\]

It can readily be verified that in terms of the above covectors one
can write 
\[
\bar{\bmg}=\epsilon_{\bmA
  \bmB}\epsilon_{\bmA'\bmB'}\bm\omega^{\bmA\bmA'}\otimes \bm\omega^{\bmB\bmB'}.
\]
Moreover, let $\{ \epsilon_\bmA{}^A \}$ denote the normalised spin
dyad giving rise to the frame $\{ \bme_{\bmA\bmA'} \}$ via the
correspondence $\epsilon_\bmA{}^A\bar{\epsilon}_{\bmA'}{}^{A'}\mapsto
e_{\bmA\bmA'}^a$. The above construction and frame will be referred in
the following discussion as the \emph{F-gauge}. Defining the spin
connection coefficients in the usual manner as
$\Gamma_{\bmA\bmA'}{}^\bmB{}_\bmC\equiv
\epsilon^\bmB{}_A\nabla_{\bmA\bmA'} \epsilon_{\bmC}{}^A $, a
computation involving the Cartan structure equations shows that the
only non-zero reduced connection coefficients are given by
\begin{align*}
&\Gamma_{\bm0\bm0'}{}^{\bm1}{}_{\bm1}=\Gamma_{\bm1\bm1'}{}^{\bm1}{}_{\bm1}
=\dfrac{\sqrt{2}}{4},
  \qquad
  \Gamma_{\bm0\bm0'}{}^{\bm0}{}_{\bm0}=\Gamma_{\bm1\bm1'}{}^{\bm0}{}_{\bm0}
=-\dfrac{\sqrt{2}}{4},
  \\ &\Gamma_{\bm1\bm0'}{}^{\bm1}{}_{\bm1}=-\Gamma_{\bm1\bm0'}{}^{\bm0}{}_{\bm0}
=\dfrac{\sqrt{2}}{4}\varpi,
  \qquad
  \Gamma_{\bm0\bm1'}{}^{\bm0}{}_{\bm0}=-\Gamma_{\bm0\bm1'}{}^{\bm1}{}_{\bm1}
=\dfrac{\sqrt{2}}{4}\overline{\varpi},
\end{align*}
were $\varpi$ is a complex function encoding the connection of
$\mathbb{S}^2$.   

\section{The massless spin-2 field equations in the F-gauge}
\label{LinearisedGravityEquations}

In this section we formulate the initial value problem, in the
F-gauge, for the spin-2 field propagating on the Minkowski spacetime
and discuss some of the basic properties of the solutions to these
equations.  The derivatives on $\mathbb{S}^2$ in these equations are
expressed in terms of the $\eth$ and $\bar\eth$ operators.

\subsection{The spin-2 equation}
\label{TheSpin2Equation}
As discussed in \cite{Val03a},
the linearised gravitational field over the Minkowski spacetime can be
described through the 
 massless spin-2 field equation
\begin{equation}\label{Spin2Equation}
\nabla_{A'}{}^{A}\phi_{ABCD}=0.
\end{equation}

\begin{remark}
{\em In the case of non-flat backgrounds this equation is overdetermined
and the spinor $\phi_{ABCD}$ is related to the Weyl curvature through
the so-called \emph{Buchdahl constraint}. This feature makes the study
of equation
\eqref{Spin2Equation} on curved backgrounds less relevant.}
\end{remark}

It is well-known that this equation is conformally invariant. It can be shown that
 equation \eqref{Spin2Equation}
implies the following  evolution equations for the components of the spinor
$\phi_{ABCD}$ 
\begin{subequations}
\begin{align}
& -(1-\tau)\bm\partial_{\tau}\phi_{0}-\rho\bm\partial_{\rho}\phi_{0}-
  \bm\partial_{+}\phi_{1} +
  \bar{\varpi}\phi_{1}=-2\phi_{0}, \label{GravEq0}\\ 
  &-\bm\partial_{\tau}\phi_{1}-\dfrac{1}{2}\bm\partial_{+}\phi_{2}-
  \dfrac{1}{2}\bm\partial_{-}\phi_{0}
  -\varpi\phi_{0}=-\phi_{1}, \label{GravEq1}\\ 
  &-\bm\partial_{\tau}\phi_{2}-\dfrac{1}{2}\bm\partial_{-}\phi_{1}-\dfrac{1}{2}
\bm\partial_{+}\phi_{3}-
\dfrac{1}{2}\varpi\phi_{1}-\dfrac{1}{2}\bar{\varpi}\phi_{3}=0,
 \label{GravEq2}  \\ 
  & -\bm\partial_{\tau}\phi_{3}-\dfrac{1}{2}\bm\partial_{+}
\phi_{4}-\dfrac{1}{2}\bm\partial_{-}\phi_{2}-
\bar{\varpi}\phi_{4}=\phi_{3}, \label{GravEq3}\\ 
& -(1+\tau)\bm\partial_{\tau}\phi_{4}+ \rho
  \bm\partial_{\rho}\phi_{4}-\bm\partial_{-}\phi_{3}+
  \varpi\phi_{3}=2\phi_{4}, \label{GravEq4}
\end{align}
\end{subequations}
 and the constraint equations
\begin{subequations}
\begin{align}
&
  \tau\bm\partial_{\tau}\phi_{1}-\rho\bm\partial_{\rho}
\phi_{1}-\dfrac{1}{2}\bm\partial_{+}\phi_{2}+
  \dfrac{1}{2}\bm\partial_{-}\phi_{0}+
  \varpi\phi_{0}=0, \label{GravEq5}
\\ &
  \tau\bm\partial_{\tau}\phi_{2}-\rho\bm
\partial_{\rho}\phi_{2}-\dfrac{1}{2}\bm\partial_{+}\phi_{3}   +
  \dfrac{1}{2}\bm\partial_{-}\phi_{1}-\dfrac{1}{2}
\bar{\varpi}\phi_{3}+
  \dfrac{1}{2}\varpi \phi_{1}=0, \label{GravEq6}
 \\ &
  \tau\bm\partial_{\tau}\phi_{3}-\rho\bm\partial_{\rho}\phi_{3}
-\dfrac{1}{2}\bm\partial_{+}\phi_{4}   +
  \dfrac{1}{2}\bm\partial_{-}\phi_{2}-\bar{\varpi}\phi_{4}=0,
 \label{GravEq7}
\end{align}
\end{subequations}
where the five components $\phi_{0},\phi_{1},\phi_{2},\phi_{3}$ 
and $\phi_{4}$, given by 
\begin{eqnarray*} 
& \phi_{0}\equiv\phi_{ABCD}o^{A}o^{B}o^{C}o^{D}, \qquad
  \phi_{1}\equiv\phi_{ABCD}o^{A}o^{B}o^{C}\iota^{D}, \\ &
  \phi_{2}\equiv\phi_{ABCD}o^{A}o^{B}\iota^{C}\iota^{D}, \qquad
  \phi_{3}\equiv\phi_{ABCD}o^{A}\iota^{B}\iota^{C}\iota^{D},\\ 
  &\phi_{4}\equiv\phi_{ABCD}\iota^{A}\iota^{B}\iota^{C}\iota^{D}& 
\end{eqnarray*}
have spin weight of $2,1,0,-1,-2$ respectively.  Here, $\{
o^{A},\iota^{A} \}$ denotes a spin dyad satisfying
\[
\tau^{AA'} = o^A \bar{o}^{A'} + \iota^A \bar{\iota}^{A'}
\]
where $\tau^{AA'}$ is the spinorial counterpart of the vector $\tau^a$
tangent to the conformal geodesics used to construct the F-gauge. It
satisfies the normalisation condition $\tau_{AA'}\tau^{AA'}=2$. The
spin dyad $\{
o^{A},\iota^{A} \}$ is defined up to a $SU(2,\mathbb{C})$
transformation. 

\medskip
Now,
one can rewrite \eqref{GravEq0}-\eqref{GravEq7} in terms of the
Newman-Penrose $\eth$ and $\bar{\eth}$ operators ---see
e.g. \cite{PenRin84,Ste91}; in particular, we make use of the
conventions used in the latter reference. A direct computation allows
to rewrite the evolution equations as
\begin{subequations}
\begin{align}
& -(1-\tau)\bm\partial_{\tau}\phi_{0}-\rho\bm\partial_{\rho}\phi_{0} +
  \eth\phi_{1}=-2\phi_{0}, \label{EvolutionGravEq0}\\ &
  -\bm\partial_{\tau}\phi_{1} + \dfrac{1}{2} \bar{\eth}\phi_{0}+
  \dfrac{1}{2}\eth\phi_{2}=-\phi_{1}, \label{EvolutionGravEq1}\\ &
  -\bm\partial_{\tau}\phi_{2} +\dfrac{1}{2}\bar{\eth}\phi_{1}+\dfrac{1}{2}
\eth \phi_{3}=0,
 \label{EvolutionGravEq2}
  \\ &
  -\bm\partial_{\tau}\phi_{3}+\dfrac{1}{2}\bar{\eth}\phi_{2}+\dfrac{1}{2}\eth\phi_{4}
 =\phi_{3}, \label{EvolutionGravEq3}\\ &
  -(1+\tau)\bm\partial_{\tau}\phi_{4}+ \rho
  \bm\partial_{\rho}\phi_{4}+ \bar{\eth} \phi_{3} = 2\phi_{4},
 \label{EvolutionGravEq4}
\end{align}
\end{subequations}
and the constraint equations in the form 
\begin{subequations}
\begin{align}
&  \tau\bm\partial_{\tau}\phi_{1}-\rho\bm\partial_{\rho}\phi_{1} + \dfrac{1}{2}
 \eth\phi_{2} - \dfrac{1}{2}\bar{\eth}\phi_{0} =0, \label{ConstraintGravEq5}
\\ &
  \tau\bm\partial_{\tau}\phi_{2}-\rho\bm\partial_{\rho}\phi_{2}
 + \dfrac{1}{2}\eth \phi_{3} - \dfrac{1}{2}\bar{\eth}\phi_{1} =0,
 \label{ConstraintGravEq6}
 \\ &
  \tau\bm\partial_{\tau}\phi_{3}-\rho\bm\partial_{\rho}\phi_{3}
+ \dfrac{1}{2}\eth \phi_{4}- \dfrac{1}{2}\bar{\eth}\phi_{2}=0. 
\label{ConstraintGravEq7}
\end{align}
\end{subequations}

\medskip 
Taking into account the spin-weight of the various components
$\phi_n$, in the following we assume that these components admit an
expansion of the form 
\beq
\phi_{n}= \sum^\infty_{\ell=|2-n|}\sum_{m=-\ell}^{\ell} \phi_{n;\ell,m}(\tau,\rho)Y_{2-n;\ell,m}, \label{eq:SphHarmDecomp}
\eeq
where $Y_{s;\ell,m}$ denotes the spin-weighted spherical harmonics
---see e.g. \cite{Ste91}. Substituting the above Ansatz into equations
\eqref{EvolutionGravEq0}-\eqref{EvolutionGravEq4} one obtains for
$\ell\geq 2$, $-\ell\leq m \leq \ell$ the
equations
\begin{subequations}
\begin{eqnarray}
&&-(1-\tau)\partial_{\tau}\phi_{0;\ell,m}-\rho\partial_{\rho}\phi_{0;\ell,m} +
  \lambda_1\phi_{1;\ell,m}+2\phi_{0;\ell,m}=0, \label{eq:spin2a}\\ 
&& - \partial_{\tau}\phi_{1;\ell,m} - \dfrac{1}{2} \lambda_1\phi_{0;\ell,m}+
  \dfrac{1}{2}\lambda_0\phi_{2;\ell,m}+\phi_{1;\ell,m}=0, \label{eq:spin2b}\\ 
  &&- \partial_{\tau}\phi_{2;\ell,m} -\dfrac{1}{2}\lambda_0\phi_{1;\ell,m}+\dfrac{1}{2}
\lambda_0\phi_{3;\ell,m}=0, \label{eq:spin2c}\\ 
&&
  -\partial_{\tau}\phi_{3;\ell,m}-\dfrac{1}{2}\lambda_0\phi_{2;\ell,m}+\dfrac{1}{2}\lambda_1\phi_{4;\ell,m}
 -\phi_{3;\ell,m}=0, \label{eq:spin2d}\\ 
 && - (1+\tau)\partial_{\tau}\phi_{4;\ell,m}+ \rho
  \partial_{\rho}\phi_{4;\ell,m}-\lambda_1 \phi_{3;\ell,m} -
    2\phi_{4;\ell,m}=0, \label{eq:spin2e} 
\end{eqnarray}
\end{subequations}
while from \eqref{ConstraintGravEq5}-\eqref{ConstraintGravEq7} one
obtains 
\begin{subequations}
\begin{eqnarray}
&&  \tau\partial_{\tau}\phi_{1;\ell,m}-\rho\partial_{\rho}\phi_{1;\ell,m} + \dfrac{1}{2}
 \lambda_0\phi_{2;\ell,m} + \dfrac{1}{2}\lambda_1\phi_{0;\ell,m} =0, \label{eq:spin2_constrainta}
\\ 
 && \tau\partial_{\tau}\phi_{2;\ell,m}-\rho\partial_{\rho}\phi_{2;\ell,m}
 + \dfrac{1}{2}\lambda_0 \phi_{3;\ell,m} + \dfrac{1}{2}\lambda_0\phi_{1;\ell,m} =0, \label{eq:spin2_constraintb}
 \\ 
  &&\tau\partial_{\tau}\phi_{3;\ell,m}-\rho\partial_{\rho;\ell,m}\phi_{3}
+ \dfrac{1}{2}\lambda_1 \phi_{4;\ell,m}+
     \dfrac{1}{2}\lambda_0\phi_{2;\ell,m}=0.  \label{eq:spin2_constraintc}
\end{eqnarray}
\end{subequations}
where
\[
\lambda_{0}
\equiv \sqrt{\ell(\ell+1)}, \qquad \lambda_{1} \equiv \sqrt{(\ell-1)(\ell+2)}.
\]

\subsection{General properties of the solutions} \label{sec:GenTheory}

In this section we provide a brief discussion of the general
properties of the solutions to the modal equations
\eqref{eq:spin2a}-\eqref{eq:spin2e} and
\eqref{eq:spin2_constrainta}-\eqref{eq:spin2_constraintc}. The basic
assumption behind the discussion in this section is:

\begin{assumption}
{\em In the following we assume that the initial data satisfies in the
  hypersurface $\mathcal{S}^\star \equiv \{\tau =0\}$
the constraint equations
\begin{align*}
&  \rho\bm\partial_{\rho}\phi_{1} - \dfrac{1}{2}
 \eth\phi_{2} + \dfrac{1}{2}\bar{\eth}\phi_{0} =0, 
\\ &
  \rho\bm\partial_{\rho}\phi_{2}
 - \dfrac{1}{2}\eth \phi_{3} +\dfrac{1}{2}\bar{\eth}\phi_{1} =0,
 \\ &
  \rho\bm\partial_{\rho}\phi_{3}
- \dfrac{1}{2}\eth \phi_{4}+\dfrac{1}{2}\bar{\eth}\phi_{2}=0., 
\end{align*}
and admit the expansion
\[
\phi_n = \sum_{p=|n-2|}^\infty \sum_{\ell=|n-2|}^p \sum_{m=-\ell}^\ell a_{n,p;\ell,m}^\star Y_{2-n;\ell,m} \rho^p.
\]
}
\end{assumption}

\begin{remark}
{\em Initial data with the above properties can be constructed using
  the methods of \cite{DaiFri01}. }
\end{remark}

We are interested in solutions which are consistent with the form of
the initial data. Thus, we consider the Ansatz:
\begin{equation}
\phi_n = \sum_{p=|n-2|}^\infty \sum_{\ell=|n-2|}^p \sum_{m=-\ell}^\ell
 a_{n,p;\ell,m}(\tau) Y_{2-n;\ell,m}  \rho^p.
\label{Solution:Ansatz}
\end{equation}

In \cite{Val03a} it has shown the following:

\begin{proposition} \label{prep.LogTermPrediction}
The coefficients of the expansion \eqref{Solution:Ansatz} satisfy:
\begin{itemize}
\item[(i)] For $\ell=0,\,1$, and all admissible $p$ and $m$ the
  coefficients $a_{n,p;\ell,m}(\tau)$ are polynomials on $\tau$.

\item[(ii)] For $p\geq 2$, $2\leq \ell \leq p-1$ and all admissible $m$,
  the coefficients $a_{n,p;\ell,m}(\tau)$ have, again, polynomial
  dependence on $\tau$.

\item[(iii)] For $p\geq 2$, $\ell =p$ and all admissible $m$, the
  coefficients $a_{n,p;p,m}(\tau)$ have logarithmic singularities
  at $\tau=\pm 1$. More precisely, the coefficients split into a part
  with polynomial (and thus smooth) dependence of $\tau$ and a
  singular part of the form
\[
(1+\tau)^{p+2-n} (1-\tau)^{p-2+n}\big( c_{n,m}
\ln(1-\tau) + d_{n,m}\ln(1+\tau)\big),
\]
with $c_{n,m}$ and $d_{n,m}$ some constant coefficients. 
The coefficients $a_{n,p;p,m}(\tau)$ are polynomial if and only if
\[
a_{0,p;p,m}^\star =a_{4,p;p,m}^\star.
\]
\end{itemize}

\end{proposition}

Finally, we notice the following result providing the link between the
\emph{formal} expansions of Ansatz \eqref{Solution:Ansatz} and actual
solutions to the spin-2 equations ---see \cite{Fri03b} for a proof:

\begin{proposition}
In a neighbourhood of $\mathcal{I}$ (including $\mathcal{I}^\pm$) the
solutions to equations
\eqref{EvolutionGravEq0}-\eqref{EvolutionGravEq4} and
\eqref{eq:spin2_constrainta}-\eqref{eq:spin2_constraintc} are of the
form
\[
\phi_n = \sum_{\ell=|n-2|}^\infty \sum_{m=-\ell}^\ell \sum_{p=|n-2|}^{N}  a_{n,p;\ell,m}(\tau) Y_{2-n;\ell,m} \rho^p + R_n, \qquad
\]
where the remainder $R_n$ is of class $C^{N-5}$.
\end{proposition}

\begin{remark}
{\em The sums in the finite expansion in the previous expression can
  be reordered at will without concerns about convergence given the
  presence of a remainder of prescribed (finite)
  differentiability. This observation justifies the \emph{formal}
  computations carried out in the formulation of our numerical scheme.}
\end{remark}

\begin{remark}
{\em Observe that the regularity of the remainder increases as one
  includes more explicit terms in the expansion. The regularity of the
explicit expansion terms is known by inspection and depends on the
particular form of the initial data.}
\end{remark}

\section{Numerical solution of the spin-2 equation near spatial infinity}\label{NumericalSolution}

In this section we exploit the general theory of the spin-2 equations
previously discussed  and rewrite the equations to obtain a
formulation which is suitable for a stable and accurate numerical
evolution.  

\medskip
We first note that the Ansatz \eqref{Solution:Ansatz} is equivalent to
\beq
\label{eq:Ansatz_phi_l}
\phi_n = \sum_{\ell=|n-2|}^\infty  \sum_{m=-\ell}^\ell \bigg( \sum_{p=\ell}^\infty 
 a_{n,p;\ell,m}(\tau)  \rho^p \bigg) Y_{2-n;\ell,m},
\eeq
i.e., we rewrite the fields $\phi_{n:l,m}$ in terms of the
spin-weighted spherical harmonics ---see eq. \eqref{eq:SphHarmDecomp}.
Moreover, we observe that
$$
\displaystyle \phi_{n:l,m}(\tau, \rho) = \sum_{p=\ell}^\infty
a_{n,p;\ell,m}(\tau)  \rho^p=\rho^\ell\sum_{i=0}^\infty
a_{n,\ell-i;\ell,m}(\tau)  \rho^i,
$$ 
which motivates the substitution
\[
\phi_{n:l,m}(\tau, \rho) = \rho^\ell C_{n;\ell,m}(\tau, \rho).
\]
The evolution equations imply
that \footnote{For simplicity in the notation, we are going to ignore
the indices ${}_{\ell,m}$ from now on.}
\begin{subequations}
\bea
&& -(1-\tau)\partial_\tau C_0 - \rho\partial_\rho C_0 + \lambda_1 C_1 + (2-\ell)C_0= 0, \label{eq:spin2_func_c0} \\
&& - \partial_\tau C_1  - \dfrac{1}{2}\lambda_1 C_0 + \dfrac{1}{2}\lambda_0 C_2 + C_1 =  0, \label{eq:spin2_func_c1}  \\
&& - \partial_\tau C_2  - \dfrac{1}{2}\lambda_0 C_1 + \dfrac{1}{2}\lambda_0 C_3 =  0, \label{eq:spin2_func_c2} \\
&& - \partial_\tau C_3  - \dfrac{1}{2}\lambda_0 C_2 + \dfrac{1}{2}\lambda_1 C_4 -C_3 =  0, \label{eq:spin2_func_c3}  \\
&& -(1+\tau)\partial_\tau C_4 + \rho\partial_\rho C_4 - \lambda_1 C_3 - (2-\ell)C_4= 0 ,\label{eq:spin2_func_c4}, 
\eea
\end{subequations}
while the constraint equations give 
\begin{subequations}
\bea
&&  \tau\partial_{\tau}C_{1} - \rho\partial_{\rho}C_{1} - \ell C_1 + \dfrac{1}{2}
 \lambda_0C_{2} + \dfrac{1}{2}\lambda_1C_{0} =0, \label{eq:spin2_constraint_func_c1}
\\ 
 && \tau\partial_{\tau}C_{2}-\rho\partial_{\rho}C_{2} - \ell C_2
 + \dfrac{1}{2}\lambda_0 C_{3} + \dfrac{1}{2}\lambda_0C_{1} =0, \label{eq:spin2_constraint_func_c2}
 \\ 
  &&\tau\partial_{\tau}C_{3}-\rho\partial_{\rho}C_{3} - \ell C_3
+ \dfrac{1}{2}\lambda_1 C_{4}+ \dfrac{1}{2}\lambda_0C_{2}=0. \label{eq:spin2_constraint_func_c3}
\eea
\end{subequations}

\begin{remark}
{\em According to Proposition \ref{prep.LogTermPrediction}, singular terms should
occur for $\ell \ge 2$. Moreover, the logarithmic terms are present
only in the term $a_{n, \ell; \ell ,m}$ from the expansion
\eqref{eq:Ansatz_phi_l}. This property indicates that we can decompose
the fields $C_{n}$ further as
\beq
\label{eq:c_Decomp}
C_{n}(\tau,\rho) = \alpha_{n}(\tau) + \rho \beta_{n}(\tau,\rho),
\eeq
with $\alpha(t)$ accounting for possible singular behaviours at
$\tau=1$ and $\beta(\tau,\rho)$ regular.}
\end{remark}

\subsection{Singular terms $\alpha(\tau)$}

As a consequence of the linearity of the field equations, the dynamics
for the coefficients $\alpha_n$  and $\beta_n$ decouples from each
other and the relevant evolution equations are obtained by
substituting the Ansatz \eqref{eq:c_Decomp} into the equations
\eqref{eq:spin2_func_c0}-\eqref{eq:spin2_constraint_func_c3}. In
particular, at $\rho=0$ we have that 
\begin{subequations}
\bea
\label{eq:EvolEq_alpha_0}
&& -(1-\tau)\partial_\tau \alpha_0
+ \lambda_1 \alpha_1 
+ (2-\ell)\alpha_0 = 0,  \\
\label{eq:EvolEq_alpha_1}
&& - \partial_\tau \alpha_1  
- \dfrac{1}{2}\lambda_1 \alpha_0 
+ \dfrac{1}{2}\lambda_0 \alpha_2
+ \alpha_1  =  0,  \\
\label{eq:EvolEq_alpha_2}
&& - \partial_\tau \alpha_2  
- \dfrac{1}{2}\lambda_0 \alpha_1 
+ \dfrac{1}{2}\lambda_0 \alpha_3=  0, 
\\
\label{eq:EvolEq_alpha_3}
&& - \partial_\tau \alpha_3  
- \dfrac{1}{2}\lambda_0 \alpha_2 
+ \dfrac{1}{2}\lambda_1 \alpha_4
-\alpha_3 =  0,  \\
\label{eq:EvolEq_alpha_4}
&& -(1+\tau)\partial_\tau \alpha_4 
- \lambda_1 \alpha_3 
- (2-\ell)\alpha_4 = 0. 
\eea
\end{subequations}
Similarly, the constraint equations imply the system
\begin{subequations}
\bea
\label{eq:ContEq_alpha_1}
&\tau\partial_\tau \alpha_1 - \ell \alpha_1 + \dfrac{1}{2}(\lambda_1\alpha_0 + \lambda_0\alpha_2)=0,  \\
\label{eq:ContEq_alpha_2}
&\tau\partial_\tau \alpha_1 -\ell \alpha_2 + \dfrac{\lambda_0}{2}(\alpha_1 + \alpha_3)=0, \\
\label{eq:ContEq_alpha_3}
&\tau\partial_\tau \alpha_1 -\ell \alpha_3 + \dfrac{1}{2}(\lambda_1\alpha_4 + \lambda_0\alpha_2)=0, 
\eea
\end{subequations}

\begin{remark}
{\em  It is
likely that this decoupling of regular and singular parts is a
property which is lost when analysing more complicated backgrounds or
the full non-linear equations. This is a question that can only be
addressed in a case-by-case basis.}
\end{remark}

At $\tau=0$, setting $\alpha^\star_n \equiv \alpha_n(0)$, equations
\eqref{eq:EvolEq_alpha_0}-\eqref{eq:EvolEq_alpha_4} and
\eqref{eq:ContEq_alpha_1}-\eqref{eq:ContEq_alpha_3} reduce to an
algebraic system, whose solution is given by
\begin{subequations}
\bea
&& \alpha^\star_1 = \dfrac{\lambda_1}{4\ell(\ell-1)}\left( \alpha^\star_0\left( 3\ell -1\right) + \alpha^\star_4\left( \ell +1\right)\right), \label{eq:ID_alpha1} \\
&& \alpha^\star_2 = \dfrac{\lambda_1\lambda_0}{2\ell(\ell-1)}\left(  \alpha^\star_0  + \alpha^\star_4 \right), \label{eq:ID_alpha2} \\
&& \alpha^\star_1 = \dfrac{\lambda_1}{4\ell(\ell-1)}\left(   \alpha^\star_0\left( \ell +1\right) + \alpha^\star_4\left( 3\ell -1\right) \right). \label{eq:ID_alpha3}
\eea
\end{subequations}
The free data in the above equations is given by $\alpha^\star_0$ and
$\alpha^\star_4$ and, according to Proposition 1 (iii), the solution
to equations \eqref{eq:EvolEq_alpha_0}-\eqref{eq:EvolEq_alpha_4}
should have polynomial dependence in $\tau$ whenever $\alpha^\star_0 =
\alpha^\star_4$. Otherwise, one obtains a logarithmic dependence of
the form 
\[
\alpha_n(\tau) \sim (1-\tau)^{\ell-2+n}\ln(1-\tau),
\]
 with the most severe singular behaviour when $\ell = 2, n=0$. 

\begin{remark}
{\em To evaluate the evolution equations numerically at $\tau=1$, we
need to calculate the fields and their first time derivatives at this
surface. Thus, we must ensure that the functions are at least of class
$C^1$ at future null infinity. An inspection of equations
\eqref{eq:EvolEq_alpha_0}-\eqref{eq:EvolEq_alpha_4} around $(1-\tau)$
reveals the behaviour
\begin{eqnarray*}
\left.
\begin{array}{c}
\alpha_0 \sim K_2\left( 1  - \dfrac{1}{2}\lambda_1^2(1-\tau) \right)
\ln(1-\tau), \vspace{0.1cm}\\
 \alpha_1 \sim \dfrac{1}{2}K_2 \lambda_1(1-\tau) \ln(1-\tau)
\end{array}
\right\} \qquad (\ell =2), \\
\alpha_0 \sim K_3 (1-\tau)\ln(1-\tau) \qquad \qquad \qquad (\ell =3), 
\end{eqnarray*}
with $K_2$ and $K_3$ constants fixed once a global solution is
obtained.} 
\end{remark}

For convenience in the numerical calculations, we introduce the auxiliary fields
\begin{subequations}
\bea
\tilde\alpha_0(\tau) &\equiv&  \alpha_0(\tau) - K \delta_{\ell,2}
\left( 1  -  \dfrac{1}{2}\lambda_1^2(1-\tau)   \right)\ln(1-\tau) \nn \\
&& - K \delta_{\ell,3}(1-\tau)\ln(1-\tau),\label{eq:AuxVar_alpha_0}  \\
\tilde\alpha_1(\tau) &\equiv&   \alpha_1(\tau) - \dfrac{1}{2}K \delta_{\ell,2} \lambda_1 (1-\tau) \ln(1-\tau), \label{eq:AuxVar_alpha_1}   \\
 \tilde\alpha_n(\tau), &\equiv&  \alpha_n(\tau) \qquad {\rm for} \quad n \ge 2, \label{eq:AuxVar_alpha_n} 
\eea
\end{subequations}
with $\delta_{\ell,\ell'}$ a Kronecker delta. In this way, $K$ picks
up the contribution of the terms proportional to the constants $K_2$
(respectively $K_3$) when $\ell =2$ (respectively $\ell=3$). For $\ell \ge 4$, we have $K=0$.

\medskip
The initial data for the the auxiliary fields $\tilde\alpha_n,$
coincides with that of $\alpha_n$ ---that is, it is given by
\eqref{eq:ID_alpha1}-\eqref{eq:ID_alpha3}. Moreover, the analogue of the left hand
side of equations \eqref{eq:EvolEq_alpha_0}-\eqref{eq:EvolEq_alpha_4}
can be readily obtained by replacing $\alpha_n$ with
$\tilde\alpha_n$. However, each one of the equations
\eqref{eq:EvolEq_alpha_0}-\eqref{eq:EvolEq_alpha_2} has, respectively,
a source term in their right hand side in the form
\begin{eqnarray*}
&& S_0 \equiv - K \delta_{\ell,2} \left( 1 - \dfrac{1}{2}\lambda_1^2(1-\tau)  \right) -   \delta_{\ell,3} K(1-\tau), \\
&& S_1 \equiv - \dfrac{1}{2}K \delta_{\ell,2}\lambda_1 \left( 1 +
   \left(1+\dfrac{\lambda_1}{2}\right)(1-\tau)\ln(1-\tau) \right) - \dfrac{1}{2} K \delta_{\ell,3}\lambda_1(1-\tau)\ln(1-\tau),   \\
&&  S_2 \equiv  \dfrac{1}{4}K \delta_{\ell,2}\lambda_0 \lambda_1(1-\tau)\ln(1-\tau).
\end{eqnarray*}

\medskip
Finally, we observe that we can straightforwardly evaluate both equation
\eqref{eq:EvolEq_alpha_0} and its time derivative at $\tau=1$ to
obtain
\begin{subequations}
\bea
&& \lambda_1 \tilde\alpha_1 + (2-\ell)\tilde\alpha_0 = - K\delta_{\ell,2}, \label{eq.alpha_tilde_0_scri} \\
&& \lambda_1 \partial_\tau\tilde \alpha_1 + (3-\ell)\partial_\tau \tilde\alpha_0 = -K \left( \dfrac{1}{2}\delta_{\ell,2}  \lambda_1^2   -  \delta_{\ell,3} \right).\label{eq.alpha_tilde_0_scri_dt}
\eea
\end{subequations}

\subsubsection{Numerical scheme}\label{sec:NumericalScheme_alpha}

We are now in the position of using spectral methods to solve the
equations numerically and inspect their
analytical behaviour. 

\medskip
In the following we fix a numerical resolution $N$ and discretise the
time coordinate $\tau\in[0,1]$ in terms the \emph{Chebyshev-Lobatto} grid points
\beq
\label{eq:taugrid_lobatto}
\tau_k \equiv  \dfrac{1}{2}\left( 1+
  \cos\left(k\dfrac{\pi}{N}\right)\right),  \qquad k=0, \ldots, N.
\eeq
The unknowns $\tilde{\alpha}_n$ will be  approximated by Chebyshev
polynomials of the first kind 
\[
T_j(x) = \cos\bigg( j\, {\arccos}\left( x\right)\bigg),
\]
so that 
\[
\tilde\alpha_n(\tau) = \sum_{j=0}^N c_{n,j} T_{j}(2\tau-1) + {\cal R}^N(\tau).
\]
The Chebyshev coefficients $c_{n,j}$ are fixed by imposing that the
residual ${\cal R}^N$ vanishes at the grid points $\tau_k$ given by 
\eqref{eq:taugrid_lobatto} ---i.e. one requires ${\cal R}^N(\tau_k) = 0$. The
derivatives of $\tilde\alpha_n$ are then calculated with the usual spectral
derivative matrices ---see e.g. \cite{CanHusQuaZan06}.

\medskip
The unknowns of this problem consist of the $5$ coefficients
$\tilde\alpha_n$ ($n=0, \ldots, 4$) evaluated at the grid points
$\tilde\alpha_n(\tau_k)$ together with one auxiliar constant $K$,
leading to a total of $n_{\rm total} = 5(N+1) + 1$ variables. The
unique solution is found once we impose the following equations at the
grid points:
\bit
\item[(i)] \textbf{Initial data at $\tau = 0$.} We provide the values
$\tilde\alpha^\star_0$ and $\tilde\alpha^\star_4$ and impose the
solution to the constraint equations
\eqref{eq:ID_alpha1}-\eqref{eq:ID_alpha2}.

\item[(ii)] \textbf{Time evolution $0< \tau \le 1$.} We impose the
evolution field equations
\eqref{eq:EvolEq_alpha_0}-\eqref{eq:EvolEq_alpha_4} written in terms
of the fields $\tilde\alpha_n$. In particular, for $n=0$, the equation
assumes the form \eqref{eq.alpha_tilde_0_scri}.

\item[(iii)] \textbf{Future null infinity $\tau =1$.} We impose the
extra condition given by the equation \eqref{eq.alpha_tilde_0_scri_dt}
if $\ell=2,3$. Otherwise, we just impose $K=0$.
\eit

This procedure leads to a linear algebraic system of $n_{\rm total}$
equations, which we solve by means of a LU decomposition.

\subsubsection{Results}
The equations are solved for two representative sets of initial data:
\begin{itemize}
\item[(i)] $\tilde\alpha^\star_0 = \tilde\alpha^\star_4= 1$;
\item[(ii)] $\tilde\alpha^\star_0 = -\tilde\alpha^\star_4/2=1$. 
\end{itemize}

\begin{figure*}[t!]
\begin{center}
\includegraphics[width=7.3cm]{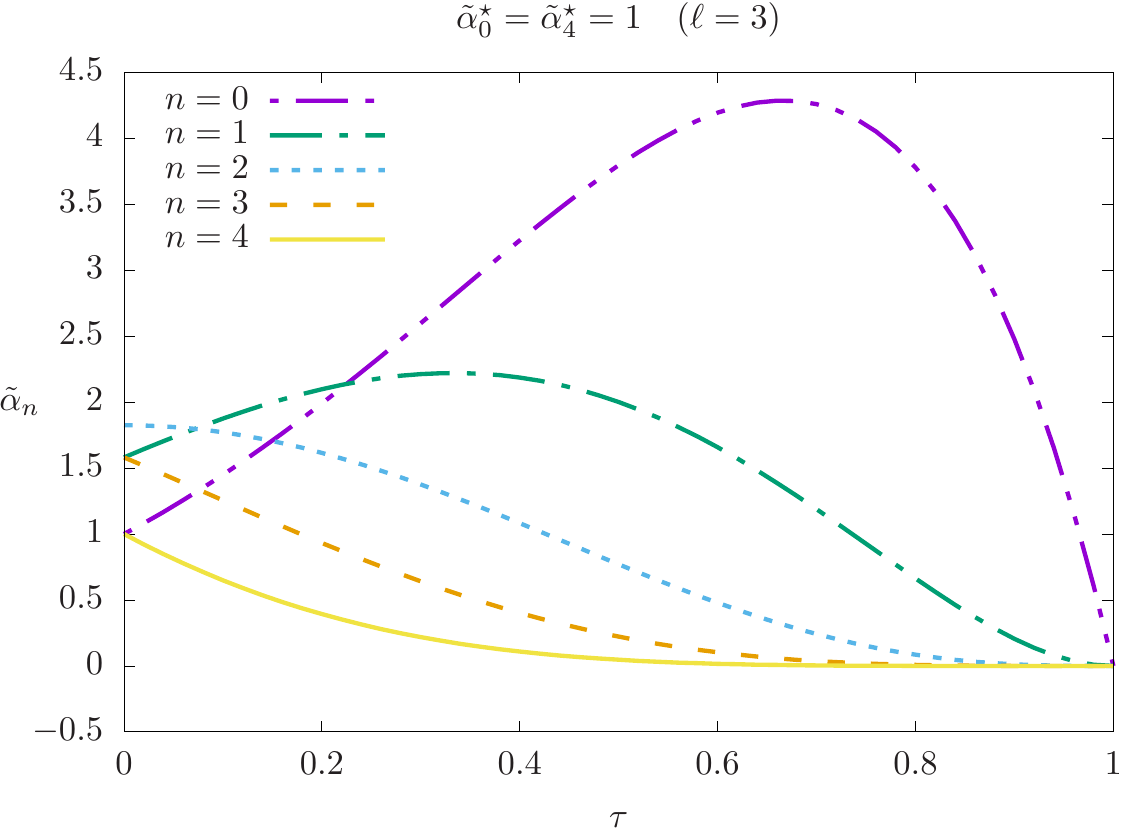}
\includegraphics[width=7.3cm]{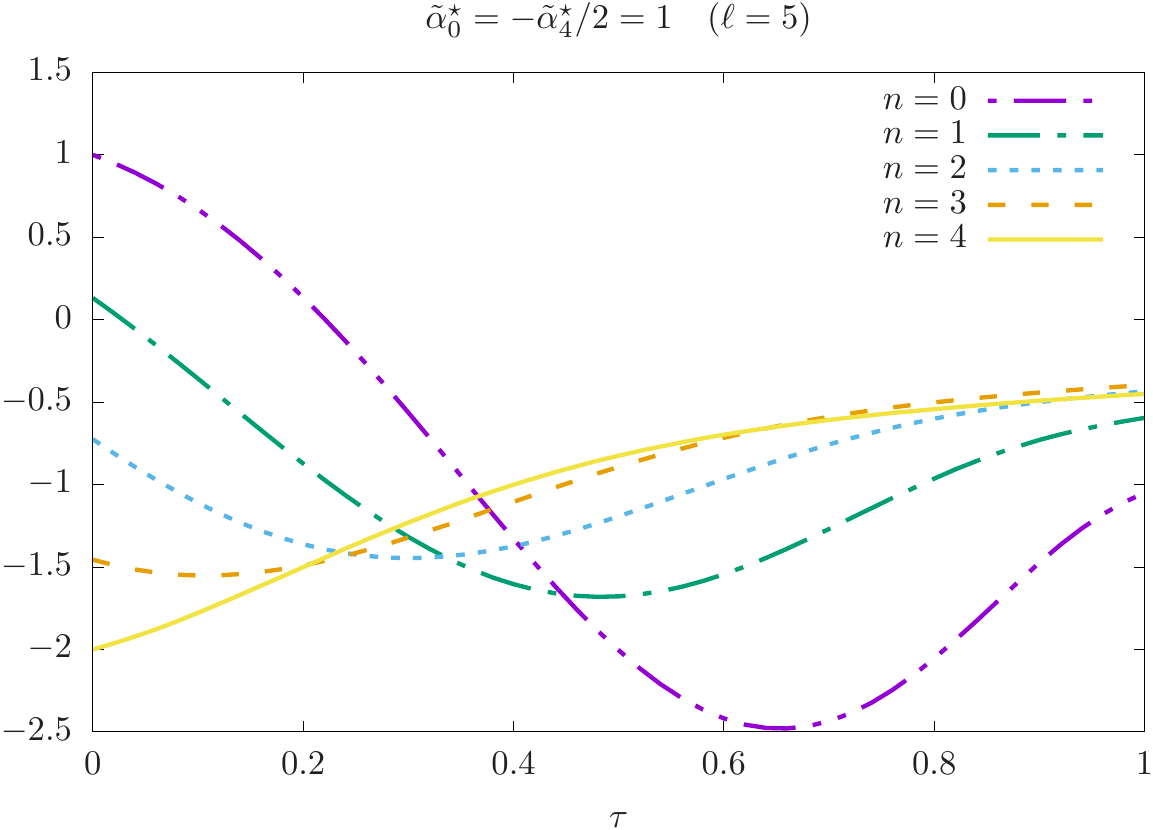}
\end{center}
\caption{Time evolution of fields $\tilde\alpha_n(\tau)$. 
According to proposition (iii), initial data with $\alpha^\star_0 = \alpha^\star_4$ (left panel)
leads to a regular evolution, whereas initial data with $\alpha^\star_0 \neq
\alpha^\star_4$ give rise to logarithmic singularities at $\tau =1$ (right panel). The singular behaviour is best appreciated in fig.~\ref{fig:cheb_alpha_log}.
}
\label{fig:alpha_evolution}
\end{figure*}

Following the general theory described in Section \ref{sec:GenTheory},
the former should lead to a polynomial dependence in time, whereas the
latter gives rise to logarithmic singularities. In Figure
\ref{fig:alpha_evolution} we provide the time evolution for this two
sets. A regular evolution resulting from initial data (i) with $\ell =
3$ is depicted in the left panel, whereas the right panel shows the
evolution of (ii) for $\ell =5$.

\begin{figure*}[b!]
\begin{center}
\includegraphics[width=7.3cm]{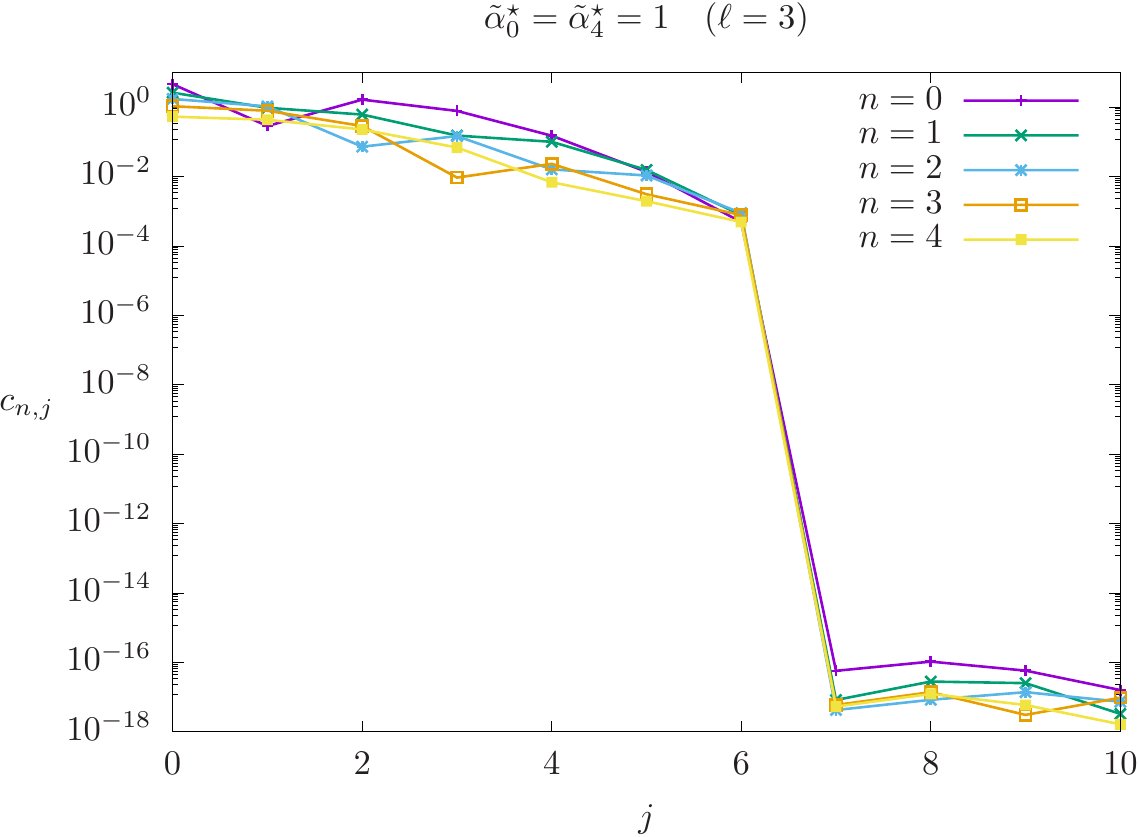}
\includegraphics[width=7.3cm]{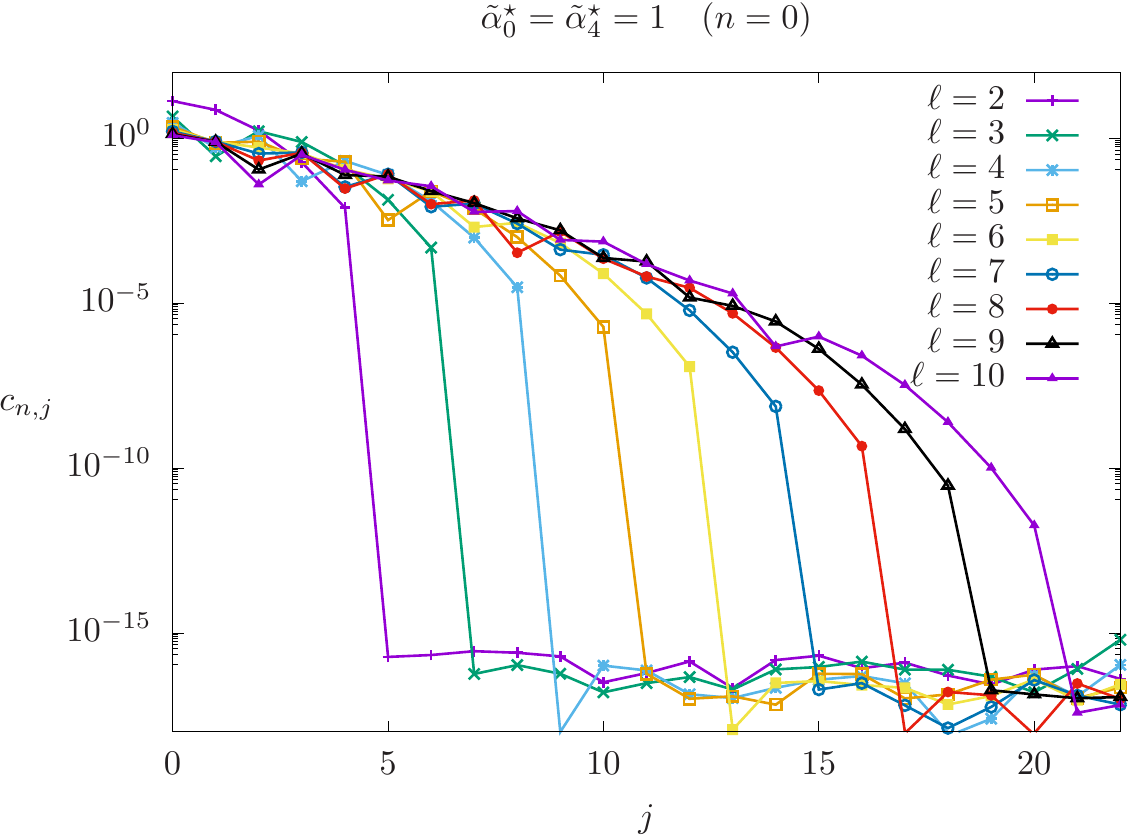}
\end{center}
\caption{Chebyshev coefficients for the polynomial time evolution of
  the 
fields $\tilde\alpha_n(\tau)$ from Figure \ref{fig:alpha_evolution} ($\ell
=3$). Left panel: the polynomial order $P$ does not depend on the
parameter $n$. Right panel: polynomial order dependence with respect
to angular parameter $\ell$. We identify $P = 2\ell$.  }
\label{fig:cheb_alpha_reg}
\end{figure*}

The behaviour of the functions $\tilde\alpha(\tau)$ is best
appreciated by studying the Chebyshev coefficients of our spectral
method. Analytic (i.e. entire) solutions display an exponential decay of the
coefficients in the form $c_j \sim \sigma^{-j}$. In particular, polynomial solutions of order $P$ are represented exactly ---in the
sense that, formally, $c_j = 0$ for $j>P$. By contrast,
singularities within the domain spoil the fast decay rate of the
Chebyshev coefficients ---and, therefore, the convergence of the
numerical error --- see \cite{CanHusQuaZan06,GraNov07} and references therein. In particular, the Chebyschev coeffcients of a function which is merely $C^k$ at
the domain boundary shows the behaviour $c_j \sim j^{-\varkappa}$,
with $\varkappa = 2k+3$.

\begin{figure*}[b!]
\begin{center}
\includegraphics[width=7.3cm]{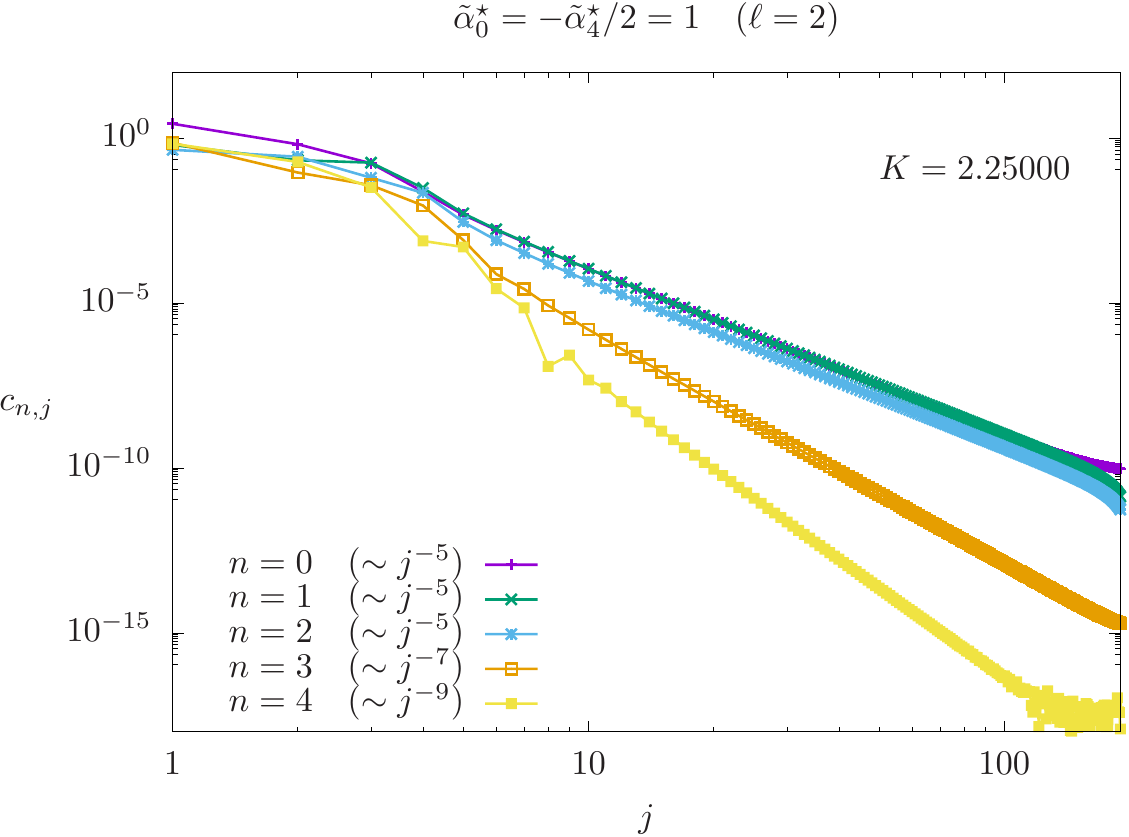}
\includegraphics[width=7.3cm]{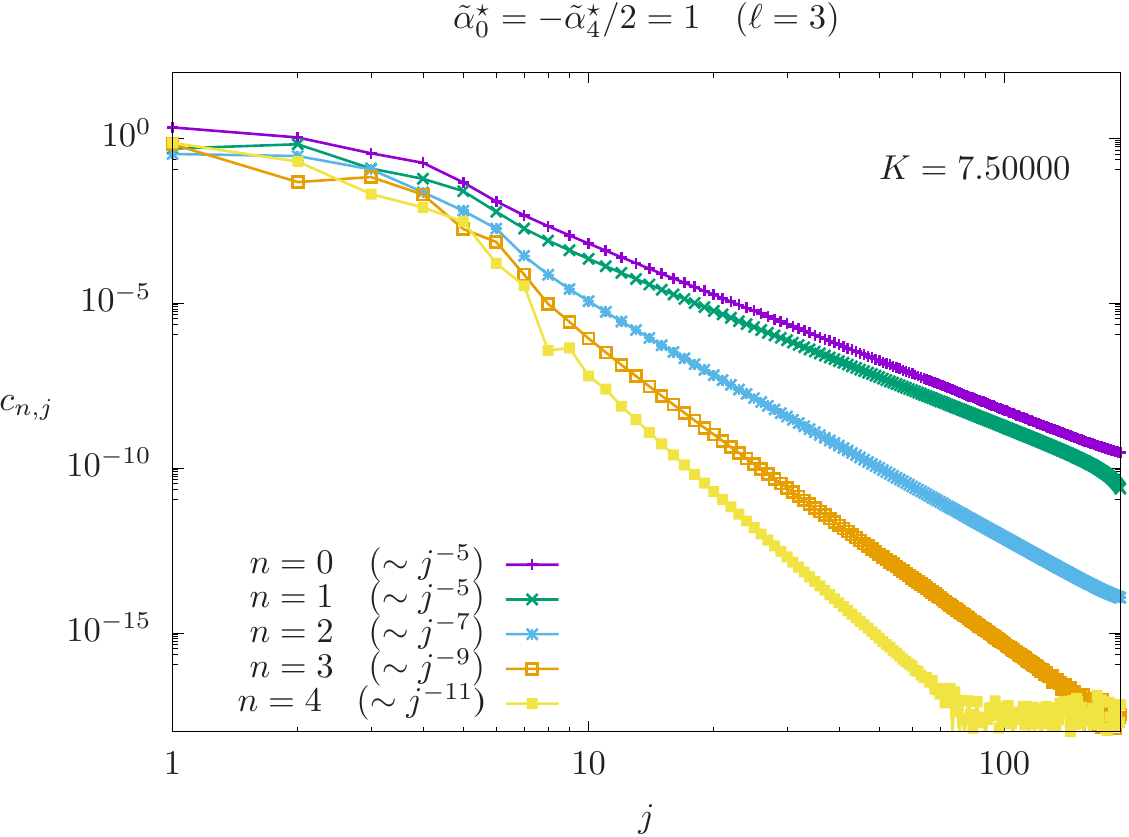}
\includegraphics[width=7.3cm]{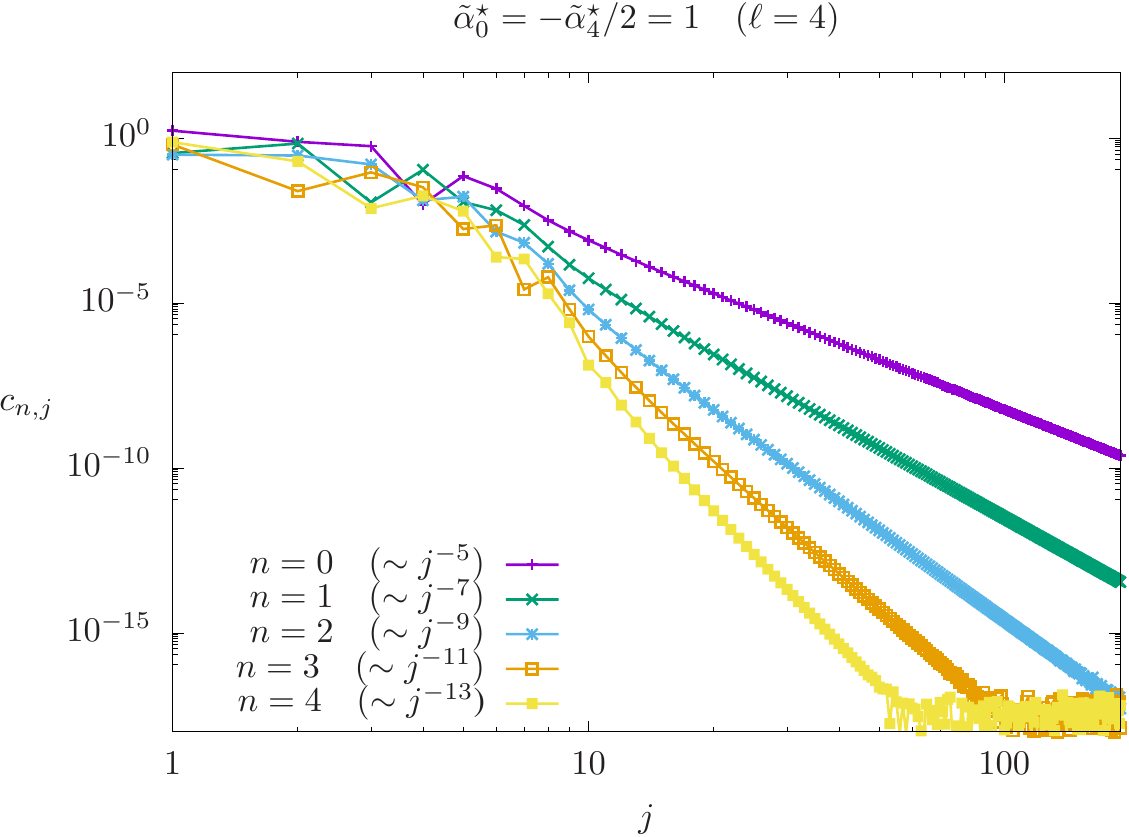}
\includegraphics[width=7.3cm]{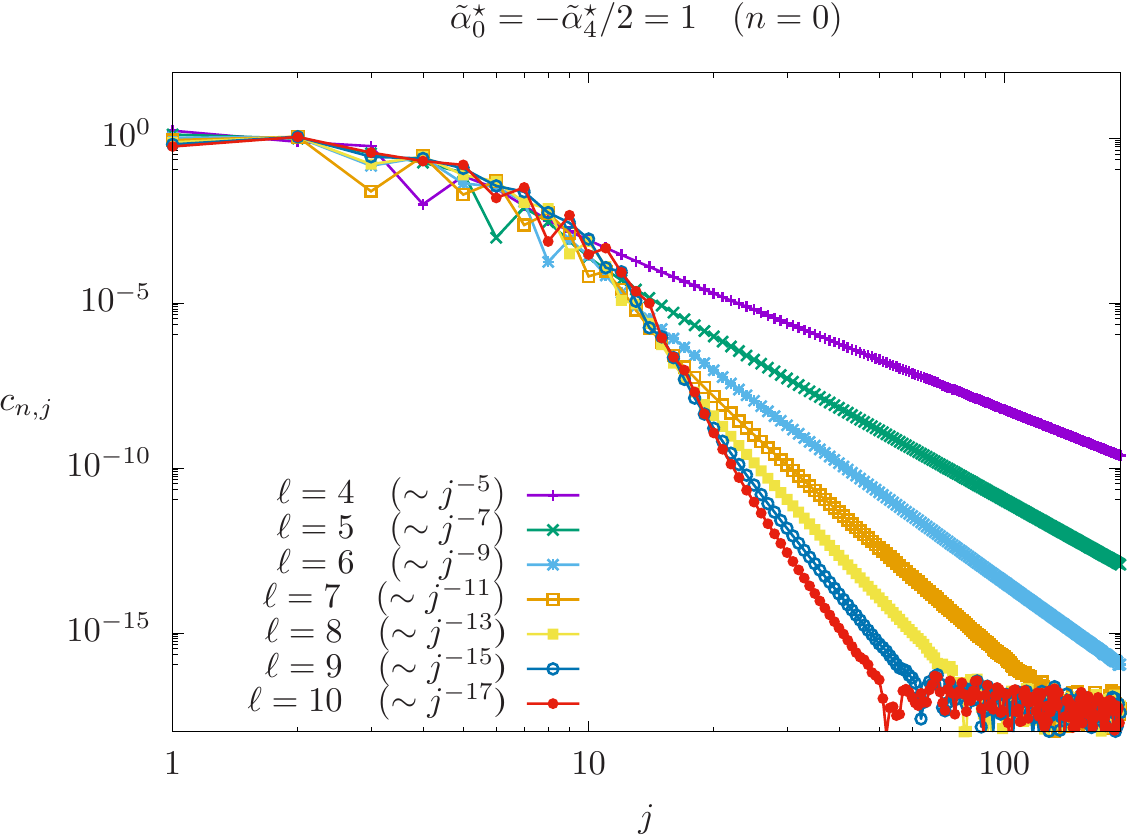}
\end{center}
\caption{Chebyshev coefficients for initial data with
$\tilde\alpha^\star_0\neq\tilde\alpha^\star_4$. The algebraic decay $c_{n,j}\sim
j^{-\varkappa}$ indicates the presence of singularities. Upper panel: the introduction of the extra variable $K$ for
$\ell=2$ (left) and $\ell=3$ (right) enforces the solution to be at
least of class $C^1$ ---the decay rate is $\varkappa \ge 5$. Lower
panel: dependence of the regularity on $n$ and $\ell$ ---fixed $\ell=4$ (left)
and fixed $n=0$ (right). The decay rate $\varkappa = 2(\ell-2+n)+1$
agrees with the theoretical discussion in Section \ref{sec:GenTheory}.}
\label{fig:cheb_alpha_log}
\end{figure*}

\medskip
\noindent
\textbf{Solutions of data of type (i).} In Figure
\ref{fig:cheb_alpha_reg}, we display the Chebyshev coefficients for
solutions arising from initial data satisfying (i). As expected,
whenever $\tilde\alpha^\star_0 = \tilde\alpha^\star_4$, we identify the polynomial
behaviour ---i.e. the value of the coefficients values drop to zero
(within the machine precision $\sim 10^{-15}$-$10^{-16}$) after a
finite number of elements. The left panel gives the coefficients for
the fields $\alpha_n$ corresponding to the evolution displayed in
Figure \ref{fig:alpha_evolution} ($\ell=3$). For a fixed value of
$\ell$, the polynomial order $P$ of the solution does not change for
the different values of $n$. Then, we compare $P$ for different values
of the parameter $\ell$. The right panel shows the results for $\ell=2
\cdots 10$ and we observe that $P = 2\ell$.

\medskip
\noindent
\textbf{Solutions of data of type (ii).} Figure
\ref{fig:cheb_alpha_log} shows the Chebyshev coefficients for the
cases $\ell = 2$ (upper-left panel) and $\ell = 3$ (upper-right
panel). The coefficients of the fields $\tilde\alpha_0$ and $\tilde\alpha_1$ for
the case $\ell = 2$ as well $\alpha_0$ with $\ell=3$ present the same
algebraic decay $\sim j^{-5}$, which agrees with the expected rate for
a $C^1$-function. This order of regularity was obtained by the
introduction of the constant $K$ accounting for the leading
logarithmic term. Its value is fixed by the algorithm as part of the
global solution of the system. Indeed, we obtain the values $K = 2.25$
($\ell =2$) and $K = 7.5$ ($\ell = 3$). In addition, we present in
Figure \ref{fig:cheb_alpha_log} the results for $\ell \ge 4$. In the
lower-left panel we fix the angular mode to $\ell=4$ and compare the
decay rate for the different $n$ values. In the lower-right panel we
concentrate on the field $n=0$ and study the decay dependence with
$\ell$. The decay rate obeys the relation $c_{n,j} \sim
j^{-\varkappa}$ with $\varkappa =2(\ell-2+n)+1$, in agreement with the
 general theory of Section \ref{sec:GenTheory}.

\subsubsection{Improving the accuracy of the calculations}

The loss of exponential convergence due to the presence of logarithmic
terms can be amended with the introduction of the coordinate
transformation
\beq
\label{eq:CoordMap_LogTerm}
\tau = 1 - \exp\left(  \dfrac{\chi}{\chi-1}\right).
\eeq
Note that $\chi\in[0,1]$ with the initial time surface still given by
$\tau=\chi=0$ and future null infinity by $\tau=\chi=1$. This change
of coordinates maps the $C^k$ functions $\tilde\alpha(\tau)$ into the
$C^\infty$ functions $\bar\alpha(\chi)\equiv
\tilde\alpha(\tau(\chi))$.  This strategy has already been implemented
on spatial coordinates in stationary problems (typically elliptic
equations) ---see
\cite{KalAns16,AmmGriJimMacMel16,KalMoeAmm17}. Recently, it has been
successfully employed along the time direction in problems dealing
with hyperbolic equations as well \cite{FraHen18}.

After rewriting equations
\eqref{eq:EvolEq_alpha_0}-\eqref{eq:EvolEq_alpha_4} in terms of the
new coordinate $\chi$, the algorithm for the numerical construction
is the same as the one outlined in Section
\ref{sec:NumericalScheme_alpha}. In particular, at $\chi=1$
equation\eqref{eq.alpha_tilde_0_scri} is still valid and we have
$\partial_\chi \bar\alpha_n = 0$ ($n=1, \ldots, 4$) as well. Moreover,
the extra condition \eqref{eq.alpha_tilde_0_scri_dt} (for $\ell=2,3$,
otherwise $K=0$) must also be taken into account. However, it is
important to first eliminate the term $\partial_\tau\tilde\alpha$ in
\eqref{eq.alpha_tilde_0_scri_dt} with the help of the evolution
equation for $\tilde\alpha_1$.

\begin{figure*}[h!]
\begin{center}
\includegraphics[width=7.3cm]{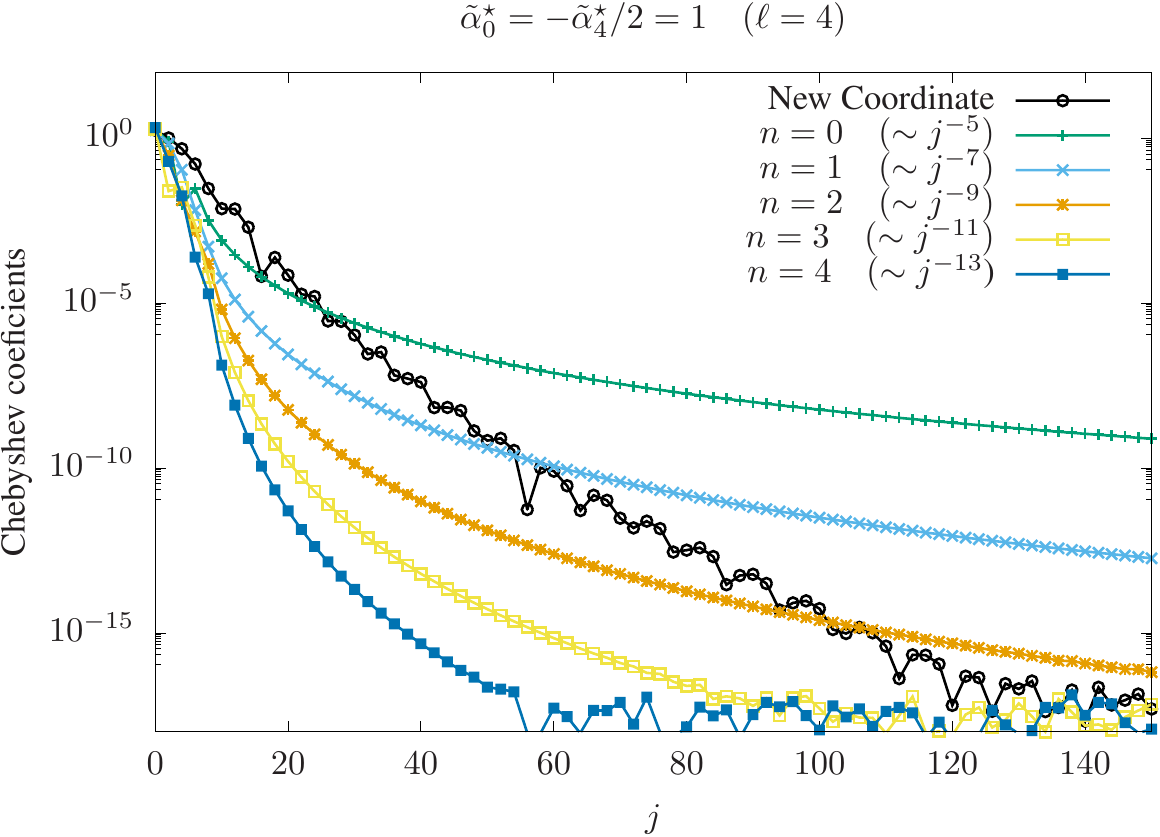}
\includegraphics[width=7.3cm]{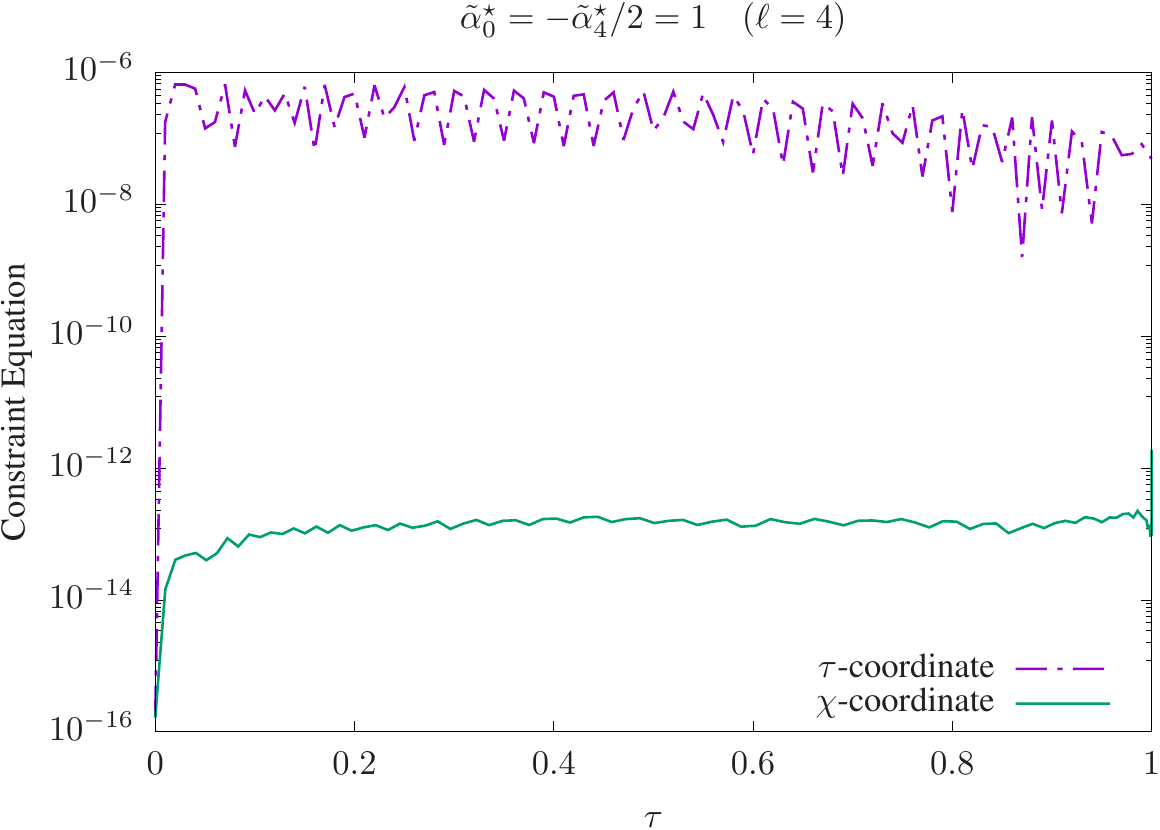}
\end{center}
\caption{Left panel: Chebyshev coefficients $\bar{c}_{n,j}$ of the
${\cal C}^\infty$ functions $\bar\alpha$ resulting from the mapping
\eqref{eq:CoordMap_LogTerm}. The decay rate is faster than algebraic
and the saturation due to round off error occurs at a resolution $\sim
100-120$. With moderate resolution, the new coordinate enhances the
accuracy of fields with an original decay rate $c_{n,j}\sim
j^{-\varkappa}$, $\varkappa \lesssim 9$. Right panel: evolution of
constraint violation ($N=100$). With the original coordinates, the
constraint equations are satisfied to order $\lesssim 10^{-6}$ due to
the logarithmic singularities. The coordinate map reduces to error to
$\sim10^{-13}$.}
\label{fig:cheb_alpha_NewCoord}
\end{figure*}

In the left panel of Figure \ref{fig:cheb_alpha_NewCoord} we compare
the Chebyshev coefficients\footnote{In Figure
\ref{fig:cheb_alpha_NewCoord} we show only $\bar{c}_{0,j}$, but the
behaviour is qualitatively the same for all $n$.} $\bar{c}_{n,j}$ of
the new functions $\bar\alpha_n(\chi)$ against the coefficients
$c_{n,j}$ of the original $\tilde\alpha_n(\tau)$. The convergence rate
for $C^\infty$-functions is still not exponential, but they decay
faster than the algebraic rate originally obtained with the
$C^k$-functions. At a moderate number of grid points, the coordinate
map \eqref{eq:CoordMap_LogTerm} is favourable for fields whose
Chebyshev coefficients decay sufficiently slow. More precisely, for a
resolution of $N \sim 80-120$, the coordinate map improves the
accuracy of the original $C^k$-functions if $k\lesssim 3$ ---the
Chebyshev coefficients of these solutions have an algebraic decay rate $c_{j}\sim
j^{-\varkappa},$ with $\varkappa \lesssim 9$. However,
for sufficiently smooth solutions (i.e. $k\gtrsim 4$), this approach makes not
much difference in practice.

Finally, we stress that the accuracy of the overall system is always limited by the
less accurate field. Nevertheless, in Figure \ref{fig:cheb_alpha_NewCoord}, we
depict in the right panel the evolution of the constraint equation
\eqref{eq:ContEq_alpha_1}. Although the constraint equations are
solved exactly at $\tau=0$, along the free evolution they are only 
satisfied up to order $\lesssim 10^{-6}$ ($N=100$) in the original
coordinates. With the change of coordinates \eqref{eq:CoordMap_LogTerm}
this error is reduced to
$\sim10^{-13}$. The same behaviour is observed for the other
constraint equations.

\subsection{Regular terms $\beta(\tau,\rho)$}

Having analysed the behaviour of the \emph{singular terms} $\alpha_n$ in the
Ansatz \eqref{eq:c_Decomp}, we now proceed to discuss the regular
terms $\beta_n$. The evolution equations for the coefficients
$\beta_n$ can be readily found to be given by
\begin{subequations}
\bea
\label{eq:EvolEq_beta_0}
&&  -(1-\tau)  \partial_\tau \beta_0 
-  \rho \partial_\rho \beta_0 
 + \lambda_1  \beta_1 
 + (1-\ell)\beta_0 = 0,  \\
\label{eq:EvolEq_beta_1}
&&- \partial_\tau \beta_1  
 - \dfrac{1}{2}\lambda_1  \beta_0
 + \dfrac{1}{2}\lambda_0  \beta_2 
 +  \beta_1 =  0,  \\
\label{eq:EvolEq_beta_2}
&&   - \partial_\tau \beta_2  
 - \dfrac{1}{2}\lambda_0  \beta_1  
 + \dfrac{1}{2}\lambda_0  \beta_3 =  0,  \\
\label{eq:EvolEq_beta_3}
&&    -  \partial_\tau \beta_3  
 - \dfrac{1}{2}\lambda_0  \beta_2
 + \dfrac{1}{2}\lambda_1  \beta_4  
-\beta_3 =  0,  \\
\label{eq:EvolEq_beta_4}
&& -(1+\tau)\partial_\tau \beta_4 
+ \rho \partial_\rho \beta_4 
 - \lambda_1  \beta_3 
 - (1-\ell) \beta_4= 0.
\eea
\end{subequations}
The associated constraint equations read
\begin{subequations}
\bea
&&\tau \partial_\tau \beta_1   -\rho\partial_{\rho}\beta_{1} - (\ell+1) \beta_1 + \dfrac{1}{2}\left(
 \lambda_0\beta_{2} + \lambda_1\beta_{0}\right) =0, \label{eq:spin2_constraint_func_beta1}
\\ 
 &&\tau \partial_\tau \beta_2 -\rho\partial_{\rho}\beta_{2} - (\ell+1) \beta_2
 + \dfrac{1}{2}\lambda_0\left( \beta_{3} + \beta_{1}\right) =0, \label{eq:spin2_constraint_func_beta2}
 \\ 
  &&\tau \partial_\tau \beta_3-\rho\partial_{\rho}\beta_{3} - (\ell+1) \beta_3
+ \dfrac{1}{2}\left( \lambda_1 \beta_{4}+ \lambda_0\beta_{2}\right)=0. \label{eq:spin2_constraint_func_beta3}
\eea
\end{subequations}

\begin{remark}
{\em
The initial data $\beta^\star_n(\rho) \equiv \beta_n(0,\rho)$ for the
evolution equations \eqref{eq:EvolEq_beta_0}-\eqref{eq:EvolEq_beta_4}
must be consistent with the constraint equations at $\tau=0$. The
procedure to construct initial data consists of specifying arbitrary
functions $\beta^\star_0(\rho)$ and $\beta^\star_4(\rho)$ and solving
equations
\eqref{eq:spin2_constraint_func_beta1}-\eqref{eq:spin2_constraint_func_beta3}
for the remaining coefficients $\beta^\star_1(\rho)$,
$\beta^\star_2(\rho)$ and $\beta^\star_3(\rho)$ in the domain
$\rho\in[0,\rho_0]$ for some convenient cut-off value $\rho_0$. The equations form a system of first order
ordinary differential equations that degenerate at
$\rho=0$. Accordingly, we impose the
natural regularity conditions at this
boundary. No other information is needed at $\rho=\rho_0$.
}
\end{remark}

The time integration of the coefficients $\beta_n$ is restricted to
equations \eqref{eq:EvolEq_beta_0}-\eqref{eq:EvolEq_beta_4}. It can be
shown that these equations form a hyperbolic system with
characteristics satisfying  
\[
\tau_{\pm}(\rho) = \pm\left( 1 - \frac{\bar{\rho}}{\rho} \right), \qquad \bar{\rho}\in[0,\rho_0].
\]
As expected, if $\bar\rho=0$ the characteristics degenerate to future
(respectively, past) null infinity $\tau = \pm 1$. Nevertheless,
following Proposition \ref{prep.LogTermPrediction}, the system of
equations should yield a regular evolution along the cylinder
$\rho=0$. In particular, at $I^+$, i.e., at the intersection of future
null infinity ($\tau=1$) and the cylinder at spatial infinity
($\rho=0$), one can eliminate all time derivatives (except for
$\partial_\tau \beta_4$) by combining the evolution equations with the
constraint equations. This procedure leads to the following conditions
at $I^+$:
\beq
\label{eq:beta_I+}
\beta_0 = \beta_4, \qquad \beta_1=\beta_3 = \sqrt{\frac{\ell-1}{\ell+2}}\beta_4 \qquad \beta_2 = \sqrt{\frac{\ell(\ell-1)}{(\ell+1)(\ell+2)}}\beta_4.
\eeq
All in all, the initial data fixes the time evolution within its future causal domain
\beq 
\label{eq:IntegrationDomain}
\bigg\{ (\tau,\rho)\in [0,1]\times[0, \rho_{\cal H}(\tau)], \quad \rho_{\cal H}(\tau) = \dfrac{\rho_0}{1+\tau}\bigg\}.
\eeq

\begin{remark}
{\em 
If one were to look for solutions within the rectangular domain
\[
\bigg\{(\tau, \rho)\in[0,1]\times[0,\rho_0]\bigg\}
\]
one would need to impose
extra boundary conditions at the regular boundary $\rho=\rho_0$. Since
we are most interested in the behaviour of the solution around
$\rho=0$, we adapt our coordinate system to the causal domain
\eqref{eq:IntegrationDomain} and avoid, in this way, the need of
imposing extra conditions at $\rho_0$. In the next section, we present
this coordinate system and discuss the numerical scheme employed to
solve the problem.}
\end{remark}

\subsubsection{Numerical scheme}
We introduce coordinates $(T, s) \in [0,1]^2$ adapted to the
integration domain \eqref{eq:IntegrationDomain} via the conditions
\[
	\tau = T, \quad	\rho = \dfrac{\rho_0}{1+T}\, s.
\] 
The constraint and evolution equations are easily re-written in terms
of $(T,s)$ and both system of equations are solved with spectral
methods. For the former, we provide a resolution $N_s$ in the radial
direction and discretise the coordinate $s$ in terms of the
Chebyshev-Lobatto grid points 
\beq
\label{eq:sgrid_lobatto}
s_i = \dfrac{1}{2}\left( 1+ \cos\left(i\dfrac{\pi}{N_s}\right)\right),  \qquad i=0\cdots N_s.
\eeq
As in Section \ref{sec:NumericalScheme_alpha}, the unknowns are approximated by
\beq
\label{eq:grid_s}
\tilde\beta^\star_n(s) = \sum_{i=0}^{N_s} c^\star_{n,i} T_{i}(2s-1) + {\cal R}^{N_s}(s),
\eeq
with coefficients $c^\star_{n,i}$ fixed by the condition ${\cal
  R}^{N_s}(s_i) = 0$. Then, we impose equations
\eqref{eq:spin2_constraint_func_beta1}-\eqref{eq:spin2_constraint_func_beta3}
at all grid points $s_i$ and the resulting linear algebraic system is
solved via a LU decomposition.

\medskip
The time evolution is performed with the fully spectral code
introduced in~\cite{MacAns14}. In addition to the radial grid \eqref{eq:grid_s}, we introduce the Chebyshev-Radau points along the time direction\footnote{While Chebyshev-Lobatto grid includes the the boundary points $s=0,1$, the Chebyshev-Radau grid incorporates only the final time slice $\tau=1$. As stated in \cite{MacAns14}, this leads to a more stable time integration in case the time direction is not compact. }
\[
T_k = \dfrac{1}{2}\left[ 1+ \cos\left(k\dfrac{2\pi}{2N_T+1}\right)\right],  \qquad k=0\cdots N_T,
\]
with $N_T$ the time resolution. The initial data is built into the
unknowns via the Ansatz
\[
\beta_n(\tau, s) = \beta^\star_n(s) + \tau\tilde\beta_n(\tau, s).
\]
Moreover, we approximate the functions $\tilde\beta_n(\tau,s)$ via
\[
\tilde\beta_n(\tau, s) = \sum_{i=0}^{N_s}\sum_{k=0}^{N_T} \tilde{c}_{n,ij} T_{i}(2s-1)T_{j}(2T-1) +{ \cal R}^{N_s,N_T}(\tau,s).
\]
As usual, the coefficients are fixed by requiring ${\cal
  R}^{N_s,N_T}(\tau_k,s_i)=0$. Equations
\eqref{eq:EvolEq_beta_0}-\eqref{eq:EvolEq_beta_4} are finally imposed
at all grid points and the system is solved with the BiConjugate Gradient 
Stabilised method, with a pre-conditioner provided by a Singly Diagonally Implicit Runge-Kutta
method~\cite{MacAns14}.

\subsubsection{Results}

\begin{figure*}[t!]
\begin{center}
\includegraphics[height=5.5cm]{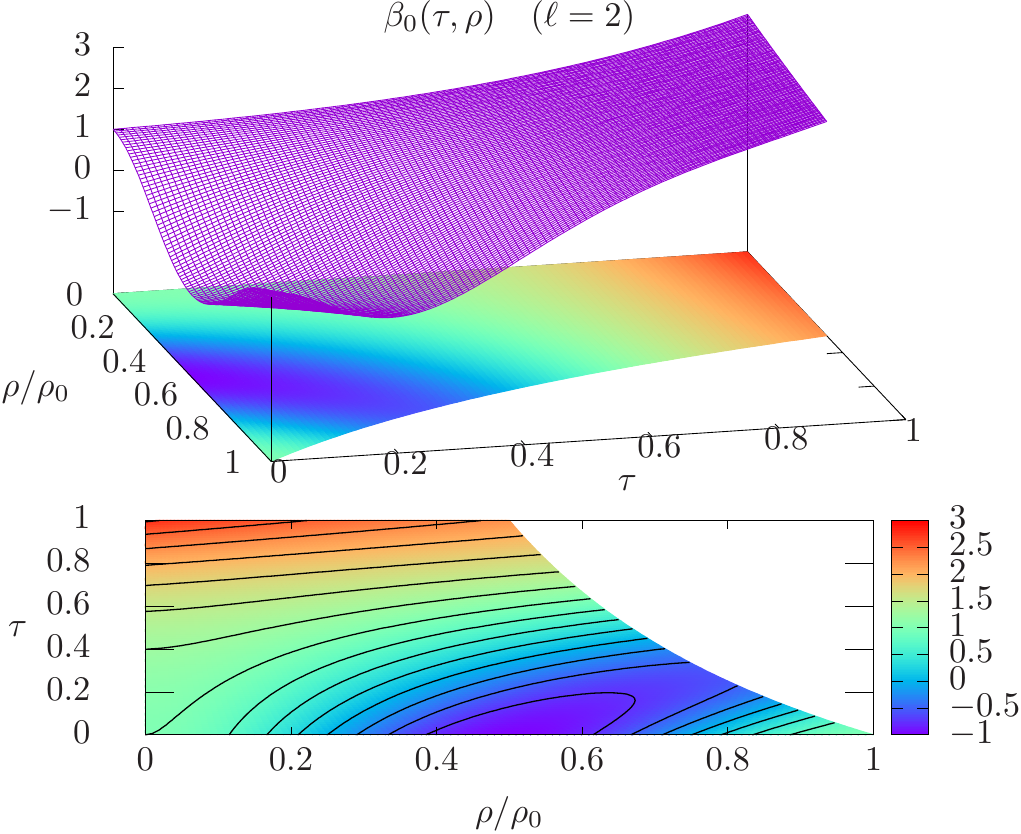}
\includegraphics[height=5.4cm]{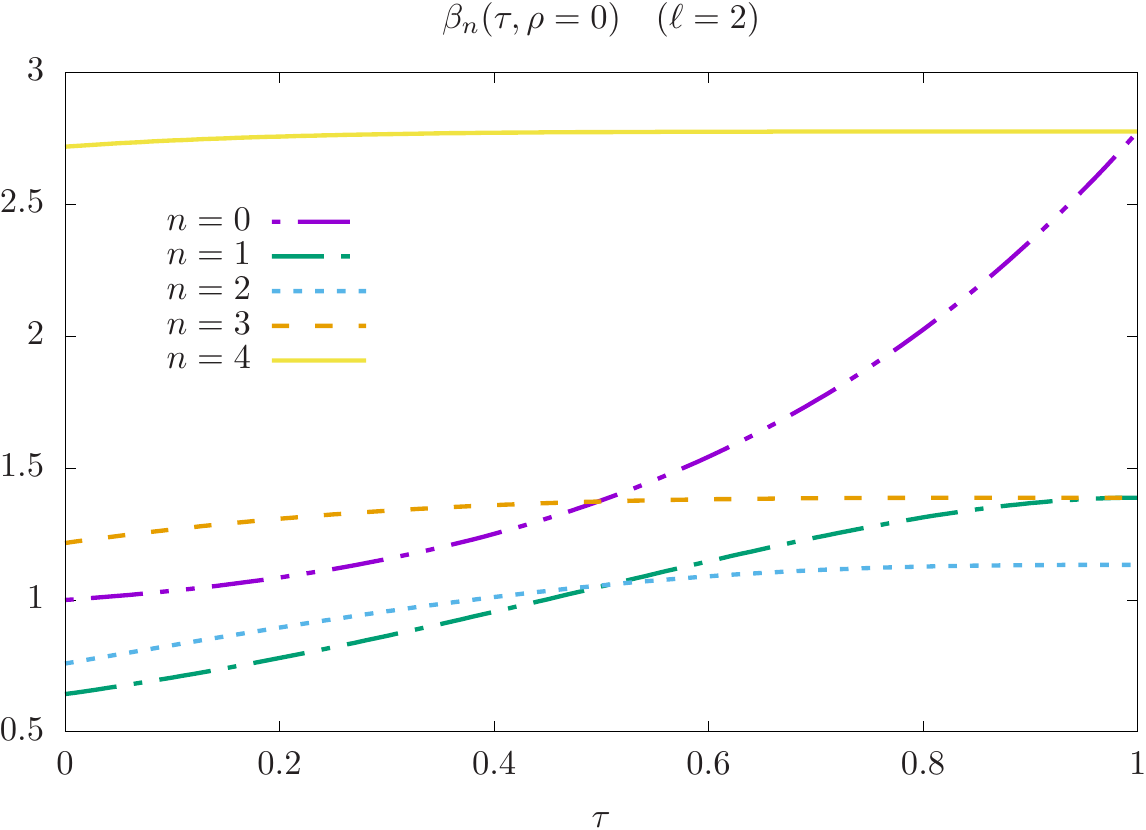}
\end{center}
\caption{Time evolution of $\beta_n(\tau,\rho)$ for $\ell=2$. Left panel: Initial data at $\tau=0$ fixes the evolution within its future causal domain \eqref{eq:IntegrationDomain}. Here $n=0$. Right panel: regular evolution along the cylinder at future null infinity ($\rho = 0$). In particular, at $I^+$ we have $\beta_0(1,0)=\beta_4(1,0)$ and $\beta_1(1,0)=\beta_3(1,0)$, as one would expect from the constraint equations.}
\label{fig:BetaSolution_l_2}
\end{figure*}

In this section, for concreteness, we discuss the properties of the
numerical solution for the field $\beta_n(\tau,\rho)$ obtained with
the resolution $N_T = 2N_s = 50$. As a particular example, we chose
the free data
\[
\beta_0^\star(\rho) = \cos(2\pi\rho), \qquad \beta_4^\star(\rho)
=e^{1-\rho}
\]
and we fix the angular parameter to $\ell=2$. Nevertheless, the
qualitative properties of the solutions are  independent of this
choice. The left panel of Figure \ref{fig:BetaSolution_l_2} shows the
evolution in time of the field $\beta_0(\tau,\rho)$ in the original
coordinates $(\tau, \rho)$ ---this choice of coordinates makes the
identification of the future causal domain of the initial data more
transparent. The right panel focuses on the time evolution along the
cylinder at $\rho=0$. As expected, at $I^+$, i.e. at $(\tau =1,
\rho=0)$, we observe $\beta_0 = \beta_4$ and $\beta_1=\beta_3$. More
specifically, the values obtained coincide with the conditions in
\eqref{eq:beta_I+}. They are a strong indication that the constraint
equations are satisfied along all the evolution. Indeed, after
checking the constraint equations we observe that they remain within
the order $\lesssim 10^{-12}$ over the whole domain of  integration.

\begin{figure*}[h!]
\begin{center}
\includegraphics[width=7.3cm]{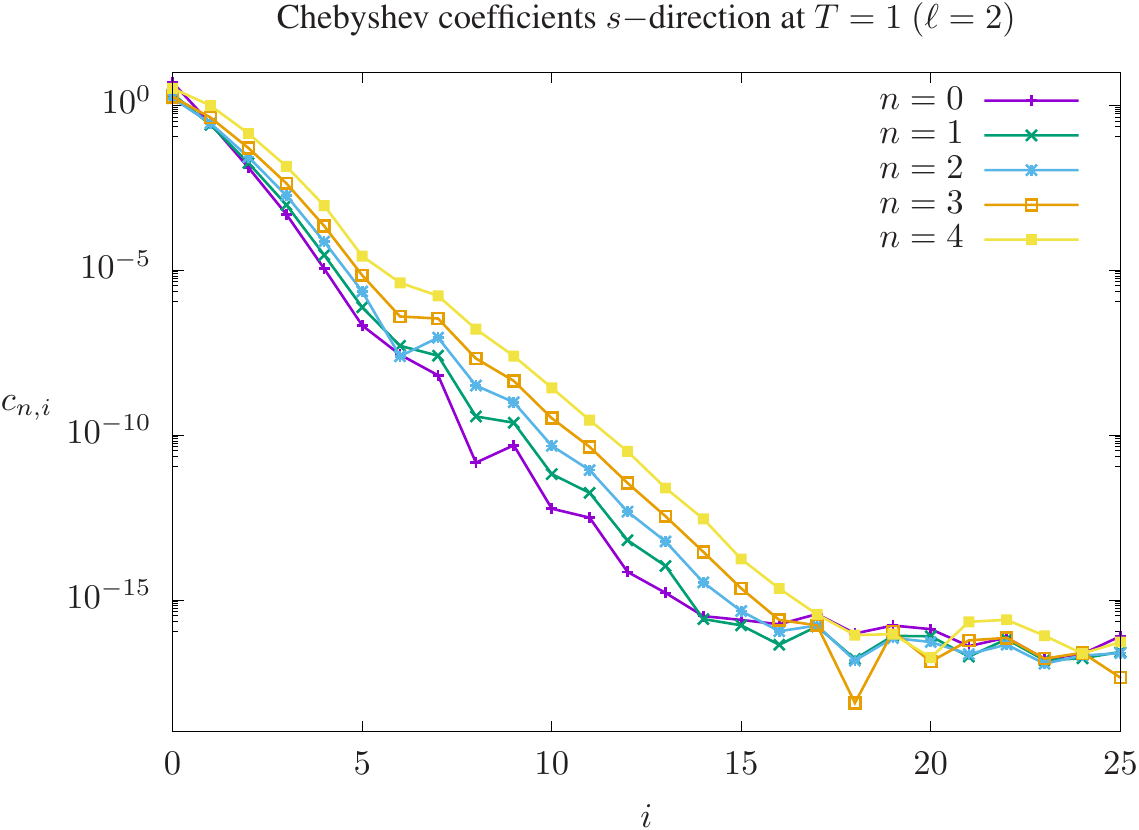}
\includegraphics[width=7.3cm]{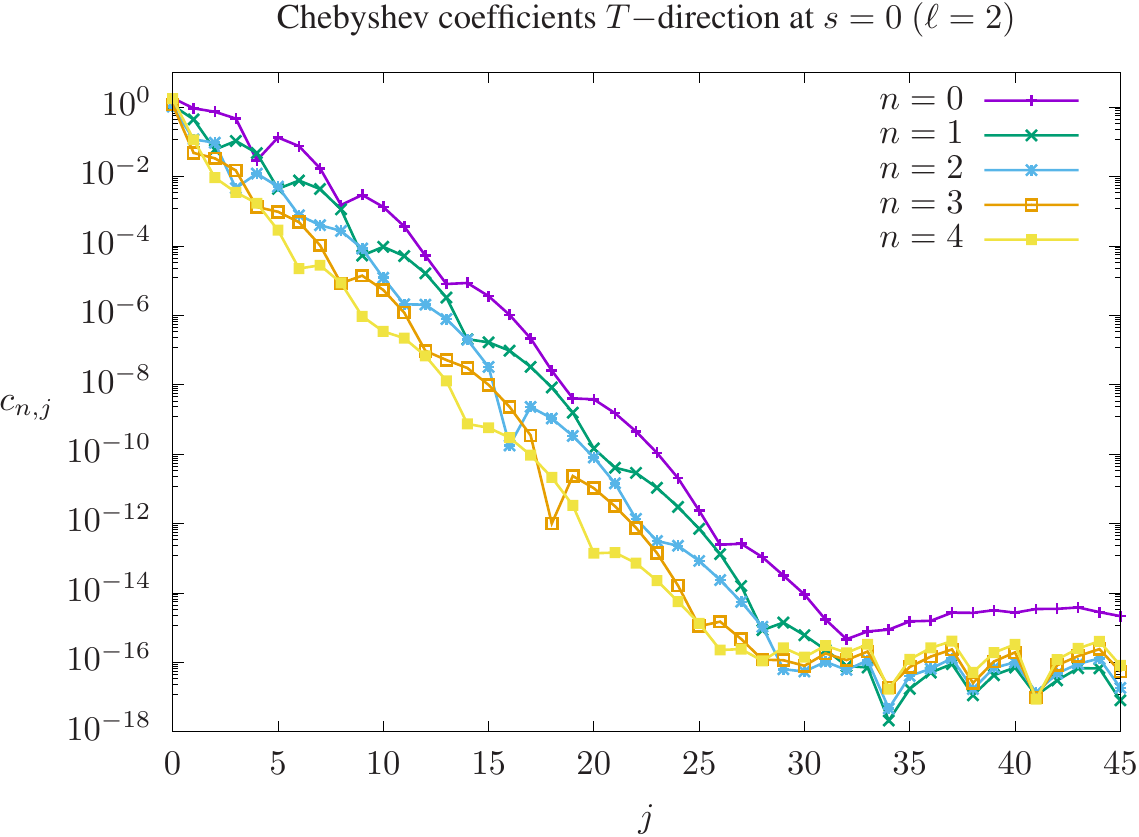}
\end{center}
\caption{Left panel: Chebyshev coefficients along the spatial direction at $T=1$. Right panel: Chebyshev coefficients along the time direction at $s=0$. In both cases we find an exponential decay, confirming the solution is regular in the integration domain.}
\label{fig:ChebBeta_l_2}
\end{figure*}

As discussed previously, we are interested in the regularity
properties of this solution. This information is obtained from the
behaviour of the Chebyshev coefficients. Particularly, we study the
analyticity of the solution with respect to both the radial and time
directions. First, we fix the time coordinate to a particular value
$T$ and study the Chebyshev coefficients $c_{n,i}$ along the
$s$-direction. We find an exponential decay for all times, indicating
that the functions are analytic with respect to the coordinate $s$
along all the time evolution. In the left panel of Figure
\ref{fig:ChebBeta_l_2} we present the results for the final time $T=1$
(future null infinity). Then, we repeat the analysis, but this time
concentrating on the Chebyshev coefficients $c_{n,j}$ along the
$T-$direction for different values of the spatial coordinate
$s$. Again, we obtain an exponential decay, in agreement with the
statements in Proposition \ref{prep.LogTermPrediction}. The right
panel of Figure \ref{fig:ChebBeta_l_2} depicts the result along $s=0$.

\section{Discussion and conclusion}\label{DiscussionConclusion}
In this article we have presented the construction of highly accurate
numerical solutions for the spin-2 equation near the cylinder at
spatial infinity of the Minkowski spacetime. The relevant field
equation can be cast as  a linear system of symmetric hyperbolic
equations subject to constraint. After decomposing the spin-2 fields
in terms of spin-weighted spherical harmonics, the equations reduce to
a system in $1+1$ dimensions, which were numerically solved with a
fully spectral code. The spectral decomposition in both the spatial
and time directions, allows to study the regularity properties of the
solutions. 

Previous analytic studies provided the necessary information to
identify the terms in the initial data whose evolution leads to
logarithm singularities. In particular, the spin-2 field was
decomposed using the Ansatz 
\[
\phi_{n;l,m}(\tau,\rho) =\rho^{\ell} \left( \alpha_{n;l,m}(\tau) +
  \rho \beta_{n;l,m}(\tau,\rho)\right), \qquad \ell\ge 2. 
\]
Thanks to the linearity of the equations, we were able to isolate each
term in order to obtain a stable and accurate numerical evolution
within the integration domain. 

The time evolution of $\alpha_{n,\ell}(\tau)$ is either polynomial or
has logarithmic singularities, depending on whether the initial data
satisfies (i) $\alpha_{0;\ell,m}(0)=\alpha_{4;\ell,m}(0)$ or (ii)
$\alpha_{0;\ell,m}(0) \neq \alpha_{4;\ell,m}(0)$, respectively. From
our numerical results it is possible to read the order $P$ for the
polynomial time evolution of initial data (i) as $P = 2\ell$.  The
logarithmic singularities present in the evolution of the initial data
of the form (ii) spoils the fast converge rate of our numerical
scheme. However, this feature allows us to study the regularity
properties of the solution. From our numerical solutions, we can
identify the presence of terms $\sim (1-\tau)^{n-2+\ell}\ln(1-\tau)$,
in agreement with the theoretical prediction. Moreover, consistent
with \cite{FraHen18}, we showed that the coordinate
mapping to deal with logarithmic terms, initially introduced to
boundary value problems, is also suitable to dynamical evolutions. 

Finally, we obtain a regular (both in space and in time) solution for
the unknowns $\beta_{n:\ell,m}(\tau,\rho)$. Note that we performed a
free evolution of the constraint system ---i.e., the constraint
equations were imposed only at the initial data surface
$\tau=0$. However, the highly accurate solution provided by the
spectral methods ensure that constraint deviations are restricted to
the machine round-off error along all the evolution.

\medskip
In future work, we plan to
extending the fully spectral code to solve the spin-2 equations
globally. A first natural step in this direction is to
follow~\cite{DouFra16} an modify the conformal
compactification of the spacetime in order to include the axis of symmetry (here,
$\rho\rightarrow \infty$). The spectral time evolution should overcome
the limitations found in previous works. Eventually, one can exploit
the use of an appropriated basis functions adapted to the topology of
the conformal time slices ---see e.g. \cite{Bey09c}--- and reduce the
evolution equations to a coupled system of ordinary differential
equations in the time coordinate, which is then solved with spectral
methods. The ultimate aim of this strategy is to depart from the
linear equations and solve the full non-linear system of Einstein
conformal field equations. As the present framework depends crucially
on the existence of a non-singular congruence of conformal geodesics
on the spacetime, the extension to the full non-linear Einstein
equations is, for the time being, restricted to situations which are
close (i.e. suitable perturbations) to \emph{background spacetimes}
which can be covered by this type of curves. Examples suitable
congruences in spherically symmetric spacetimes have been studied in
e.g. \cite{Fri03c,LueVal13b,GarGasVal18}. For more complicated spacetimes
(e.g. the Kerr solution) the formation of caustics is a possibility.

\subsection*{Acknowledgements}
This work was partially supported by the the European Research Council
Grant No. ERC-2014-StG 639022-NewNGR ``New frontiers in numerical
general relativity". We acknowledge that the results of this research
have been achieved using the DECI resource Cartesius based in
Netherlands with support from the PRACE aisbl DECI-14 14DECI0017
NRBA. R.P.M. thankfully acknowledge the computer resources at
MareNostrum and the technical support provided by Barcelona
Supercomputing Center (FI-2017-3-0012 ``New frontiers in numerical
general relativity").



\begin{thebibliography}{10}

\bibitem{AceVal11}
A.~{Ace\~{n}a} \& J.~A. {Valiente Kroon},
\newblock {\em Conformal extensions for stationary spacetimes},
\newblock Class. Quantum Grav. {\bf 28}, 225023 (2011).

\bibitem{AmmGriJimMacMel16}
M.~Ammon, S.~Grieninger, A.~Jimenez-Alba, R.~Macedo, \& L.~Melgar,
\newblock {\em Holographic quenches and anomalous transport},
\newblock JHEP {\bf 1609}, 131 (2016).

\bibitem{AnsHen11}
  M.~Ansorg \& J.~Hennig,
  \newblock{ \em The Interior of axisymmetric and stationary black holes: Numerical and analytical studies},
 \newblock  J.\ Phys.\ Conf.\ Ser.\  {\bf 314} (2011) 012017

\bibitem{Ash80}
A.~Ashtekar,
\newblock {\em Asymptotic structure of the gravitational field at spatial
  infinity},
\newblock in {\em General Relativity and Gravitation: One hundred years after
  the birth of Albert Einstein}, edited by A.~Held, volume~2, page~37, Plenum
  Press, 1980.

\bibitem{AshHan78}
A.~Ashtekar \& R.~O. Hansen,
\newblock {\em A unified treatment of null and spatial infinity in general
  relativity. {I}. {U}niversal structure, asymptotic symmetries, and conserved
  quantities at spatial infinity},
\newblock J. Math. Phys. {\bf 19}, 1542 (1978).

\bibitem{Bei84}
R.~Beig,
\newblock {\em Integration of {Einstein}'s equations near spatial infinity},
\newblock Proc. Roy. Soc. Lond. A {\bf 391}, 295 (1984).

\bibitem{BeiSch82}
R.~Beig \& B.~G. Schmidt,
\newblock {\em {Einstein}'s equation near spatial infinity},
\newblock Comm. Math. Phys. {\bf 87}, 65 (1982).

\bibitem{Bey09c}
F.~Beyer,
\newblock {\em A spectral solver for evolution problems with spatial
  S3-topology},
\newblock J. Comput. Phys. {\bf 228}, 6496 (2009).

\bibitem{BeyDouFraWha12}
F.~Beyer, G.~Doulis, J.~Frauendiener, \& B.~Whale,
\newblock {\em Numerical space-times near space-like and null infinity. The
  spin-2 system on Minkowski space},
\newblock Class. Quantum Grav. {\bf 29}, 245013 (2012).

\bibitem{BeyDouFraWha13}
F.~Beyer, G.~Doulis, J.~Frauendiener, \& B.~Whale,
\newblock {\em The Spin-2 Equation on Minkowski Background},
\newblock Springer Proc. Math. Stat. {\bf 60}, 465 (2014).

\bibitem{CanHusQuaZan06}
C.~Canuto, M.~Hussaini, A.~Quarteroni, \& T.~Zang,
\newblock {\em Spectral Methods: Fundamentals in Single Domains},
\newblock Springer, 2006.

\bibitem{DaiFri01}
S.~Dain \& H.~Friedrich,
\newblock {\em Asymptotically flat initial data with prescribed regularity at
  infinity},
\newblock Comm. Math. Phys. {\bf 222}, 569 (2001).

\bibitem{DouFra13}
G.~Doulis \& J.~Frauendiener,
\newblock {\em The second order spin-2 system in flat space near space-like and
  null-infinity},
\newblock Gen. Rel. Grav. {\bf 45}, 1365 (2013).

\bibitem{DouFra16}
G.~Doulis \& J.~Frauendiener,
\newblock {\em Global simulations of Minkowski spacetime including spacelike
  infinity},
\newblock Phys. Rev. D {\bf 95}, 024035 (2017).

\bibitem{FraHen14}
J.~Frauendiener \& J.~Hennig,
\newblock {\em Fully pseudospectral solution of the conformally invariant wave
  equation near the cylinder at spacelike infinity},
\newblock Class. Quantum Grav. {\bf 31}, 085010 (2014).

\bibitem{FraHen17}
J.~Frauendiener \& J.~Hennig,
\newblock {\em Fully pseudospectral solution of the conformally invariant wave
  equation near the cylinder at spacelike infinity. II: Schwarzschild
  background},
\newblock Class. Quantum Grav. {\bf 34}, 045005 (2017).

\bibitem{FraHen18}
J.~Frauendiener \& J.~Hennig,
\newblock {\em Fully pseudospectral solution of the conformally invariant wave
  equation near the cylinder at spacelike infinity. III: Nonspherical
  Schwarzschild waves and singularities at null infinity},
\newblock Class. Quantum Grav. {\bf 35}, 065015 (2018).

\bibitem{Fri81a}
H.~Friedrich,
\newblock {\em On the regular and the asymptotic characteristic initial value
  problem for {Einstein}'s vacuum field equations},
\newblock Proc. Roy. Soc. Lond. A {\bf 375}, 169 (1981).

\bibitem{Fri84}
H.~Friedrich,
\newblock {\em Some (con-)formal properties of {Einstein}'s field equations and
  consequences},
\newblock in {\em Asymptotic behaviour of mass and spacetime geometry.
  {L}ecture notes in physics 202}, edited by F.~J. Flaherty, Springer Verlag,
  1984.

\bibitem{Fri86b}
H.~Friedrich,
\newblock {\em On the existence of n-geodesically complete or future complete
  solutions of {E}instein's field equations with smooth asymptotic structure},
\newblock Comm. Math. Phys. {\bf 107}, 587 (1986).

\bibitem{Fri95}
H.~Friedrich,
\newblock {\em {Einstein} equations and conformal structure: existence of
  anti-de {Sitter}-type space-times},
\newblock J. Geom. Phys. {\bf 17}, 125 (1995).

\bibitem{Fri98a}
H.~Friedrich,
\newblock {\em Gravitational fields near space-like and null infinity},
\newblock J. Geom. Phys. {\bf 24}, 83 (1998).

\bibitem{Fri03a}
H.~Friedrich,
\newblock {\em Conformal Einstein evolution},
\newblock in {\em The conformal structure of spacetime: Geometry, Analysis,
  Numerics}, edited by J.~Frauendiener \& H.~Friedrich, Lecture Notes in
  Physics, page~1, Springer, 2002.

\bibitem{Fri03c}
H.~Friedrich,
\newblock {\em Conformal geodesics on vacuum spacetimes},
\newblock Comm. Math. Phys. {\bf 235}, 513 (2003).

\bibitem{Fri03b}
H.~Friedrich,
\newblock {\em Spin-2 fields on Minkowski space near space-like and null
  infinity},
\newblock Class. Quantum Grav. {\bf 20}, 101 (2003).

\bibitem{FriKan00}
H.~Friedrich \& J.~K\'{a}nn\'{a}r,
\newblock {\em Bondi-type systems near space-like infinity and the calculation
  of the {N}{P}-constants},
\newblock J. Math. Phys. {\bf 41}, 2195 (2000).

\bibitem{GasVal16}
E.~Gasper\'{\i}n \& J.~A. {Valiente Kroon},
\newblock {\em Zero rest-mass fields and the Newman-Penrose constants on flat
  space},
\newblock Available at {\tt arXiv: 1608.05716[gr-qc]}.

\bibitem{GarGasVal18}
A. Garc\'{\i}a-Parrado, E.~Gasper\'{\i}n \& J.~A. {Valiente Kroon},
\newblock {\em Conformal geodesics in the Schwarzshild-de Sitter and Schwarzschild anti-de Sitter spacetimes},
\newblock Class. Quantum Grav. {\bf 35}, 045002 (2018).

\bibitem{GraNov07}
P.~Grandclement \& J.~Novak,
\newblock {\em Spectral methods for numerical relativity},
\newblock Living Rev. Rel. {\bf 12}, 1 (2007).

\bibitem{HenAns09}
J.~Hennig \& M.~Ansorg,
\newblock {\em A fully pseudospectral scheme for solving singular hyperbolic
  equations},
\newblock J. Hyp. Diff. Eqns. {\bf 6}, 161 (2009).


\bibitem{Hen12}
  J.~Hennig,
  \newblock {\em Fully pseudospectral time evolution and its application to 1+1 dimensional physical problems},
  \newblock J.\ Comput.\ Phys.\  {\bf 235} (2013) 322

\bibitem{Hub99a}
P.~H\"{u}bner,
\newblock {\em How to avoid artificial boundaries in the numerical calculation
  of black hole spacetimes},
\newblock Class. Quantum Grav. {\bf 16}, 2145 (1999).

\bibitem{Hub99b}
P.~H\"{u}bner,
\newblock {\em A scheme to numerically evolve data for the conformal {Einstein}
  equation},
\newblock Class. Quantum Grav. {\bf 16}, 2823 (1999).

\bibitem{Hub01b}
P.~H\"{u}bner,
\newblock {\em From now to timelike infinity on a finite grid},
\newblock Class. Quantum Grav. {\bf 18}, 1871 (2001).

\bibitem{KalAns16}
M.~Kalisch \& M.~Ansorg,
\newblock {\em Pseudo-spectral construction of non-uniform black string
  solutions in five and six spacetime dimensions},
\newblock Class. Quantum Grav. {\bf 33}, 215005 (2016).

\bibitem{KalMoeAmm17}
M.~Kalisch, S.~Moeckel, \& M.~Ammon,
\newblock {\em Critical behavior of the black hole/black string transition},
\newblock JHEP {\bf 1708}, 49 (2017).

\bibitem{LueVal09}
C.~L\"ubbe \& J.~A. {Valiente Kroon},
\newblock {\em On de Sitter-like and Minkowski-like spacetimes},
\newblock Class. Quantum Grav. {\bf 26}, 145012 (2009).

\bibitem{LueVal13b}
C.~L\"ubbe \& J.~A. {Valiente Kroon},
\newblock {\em A class of conformal curves in the Reissner-Nordström spacetime},
\newblock Ann. H. Poincar\'e. {\bf 15}, 1327 (2013).

\bibitem{MacAns14}
R.~P. Macedo \& M.~Ansorg,
\newblock {\em Axisymmetric fully spectral code for hyperbolic equations},
\newblock J. Comput. Phys. {\bf 276}, 357 (2014).

\bibitem{PenRin84}
R.~Penrose \& W.~Rindler,
\newblock {\em Spinors and space-time. {V}olume 1. {T}wo-spinor calculus and
  relativistic fields},
\newblock Cambridge University Press, 1984.

\bibitem{SanSop18}
  D.~Santos-Oliv\'{a}n and C.~F.~Sopuerta,
\newblock {\em Pseudo-Spectral Collocation Methods for Hyperbolic Equations with Arbitrary Precision: Applications to Relativistic Gravitation},
\newblock arXiv:1803.00858 [physics.comp-ph].

\bibitem{Ste91}
J.~Stewart,
\newblock {\em Advanced general relativity},
\newblock Cambridge University Press, 1991.

\bibitem{Tod02}
K.~P. Tod,
\newblock {\em Isotropic cosmological singularities},
\newblock in {\em The Conformal structure of space-time. Geometry, Analysis,
  Numerics}, edited by J.~Frauendiener \& H.~Friedrich, Lect. Notes. Phys.
  \textbf{604}, page 123, 2002.

\bibitem{Val03a}
J.~A. Valiente~Kroon,
\newblock {\em Polyhomogeneous expansions close to null and spatial infinity},
\newblock in {\em The Conformal Structure of Spacetimes: Geometry, Numerics,
  Analysis}, edited by J.~Frauendiner \& H.~Friedrich, Lecture Notes in
  Physics, page 135, Springer, 2002.

\bibitem{CFEBook}
J.~A. {Valiente Kroon},
\newblock {\em Conformal Methods in General Relativity},
\newblock Cambridge University Press, 2016.

\end{thebibliography}

\end{document}